\documentstyle[preprint,aps,epsf]{revtex}
\newcommand{\del}{\mbox{$\Delta$}}
\newcommand{\arr}{\mbox{$\longrightarrow$}}
\newcommand{\pinphin}{\mbox{$\pi N \arr \phi N$}}
\newcommand{\pidelnphi}{\mbox{$\pi \del \arr \phi N$}}
\newcommand{\nnnnphi}{\mbox{$N N \arr N N \phi$}}
\newcommand{\ndelnnphi}{\mbox{$N \del \arr N N \phi$}}
\newcommand{\deldelnnphi}{\mbox{$\del \del \arr N N \phi$}}
\newcommand{\kkphi}{\mbox{$K \bar K \arr \phi$}}
\newcommand{\pirhophi}{\mbox{$\pi \rho \phi$}}
\newcommand{\phinpin}{\mbox{$\phi N \arr \pi N$}}
\newcommand{\phinpiDel}{\mbox{$\phi N \arr \pi \Delta$}}
\newcommand{\phinnrho}{\mbox{$\phi N \arr \rho N$}}

\newcommand{\phinlambdak}{\mbox{$\phi N \arr K\Lambda $}}

\newcommand{\ppppphi}{\mbox{$p p \arr p p \phi$}}
\begin{document}

\title{Phi Meson Production in Heavy-Ion Collisions at SIS Energies}

\author{W. S. Chung$^{1}$,  G. Q. Li$^{1,2}$, and C. M. Ko$^{1}$}

\address{$^{1}$Department of Physics and Cyclotron Institute,
Texas A\&M University, \\
College Station, Texas 77843, U.S.A.\\
$^{2}$Department of Physics, State University of New York at Stony Brook,\\
Stony Brook, New York 11794, U.S.A.}

\maketitle

\begin{abstract}
Phi meson production in heavy-ion collisions at SIS/GSI energies 
($\sim 2$ GeV/nucleon) is studied in the relativistic transport model.
We include contributions from baryon-baryon, pion-baryon, and 
kaon-antikaon collisions. The cross sections for the first two processes 
are obtained in an one-boson-exchange model, while that for the last process
is taken to be of Breit-Wigner form through the phi meson resonance.
The dominant contribution to phi meson production in heavy ion collisions
at these energies is found to come from secondary pion-nucleon collisions. 
Effects due to medium modifications of kaon masses are also studied and 
are found to reduce the phi meson yield by about a factor of two, mainly 
because of increased phi decay width as a result of dropping kaon-antikaon 
masses. In this case, the $\phi/K^-$ ratio is about 4\%, which is a 
factor of 2-3 below preliminary experimental data from the FOPI 
collaboration at GSI.  Including also the reduction of phi meson mass 
in medium increases this ratio to about 8\%, which is then in reasonable 
agreement with the data.

\end{abstract}

\newpage

\section{Introduction}
\label{intro}
There are increasing interests in phi meson production from heavy ion 
collisions since the suggestion by Shor \cite{shor} that an enhanced 
production may signal the formation of a quark-gluon plasma
in these collisions as its production in hadronic matter is usually 
suppressed by the OZI rule \cite{ozi}. However, recent
theoretical studies have suggested that because of chiral symmetry 
restoration hadron properties are expected to change in hot dense hadronic 
matter \cite{brown,hat,asa} (For recent reviews see 
Refs. \cite{brownrho,mfrev,kkl}). For phi meson in dense matter, QCD 
sum rule calculations have shown that its mass decreases by about 2-3\% 
at normal nuclear matter density \cite{hat}. Reduced in-medium phi meson
mass has also been predicted in vector dominance model with dropping kaon 
masses \cite{ko92}, in hadronic models that include the particle-antiparticle 
vacuum polarization \cite{hat95}, and in hidden gauge theory
\cite{song}. A reduced phi meson mass in medium is expected to enhance
its yield from heavy ion collisions even in the absence of the 
quark-gluon plasma \cite{kosa}. In this case, one can also study the
phi meson in-medium properties via the dilepton spectrum from these 
collisions as pointed out in Refs. \cite{hat,karsch,kkonly}.
At high temperature, the reduction of phi meson mass may 
become more appreciable because of the decrease of 	
strangeness condensate $\langle{\bar s}s\rangle$ as a result of the 
presence of strange hadrons \cite{asa}. It has thus been suggested that
a low mass phi peak besides the normal one may appear in the dilepton
spectrum from ultrarelativistic heavy ion collisions if a first-order
phase transition or a slow cross-over between the quark-gluon plasma 
and the hadronic matter occurs in the collisions \cite{ak94}.

Experimentally, phi meson production has been
studied at SPS/CERN by the NA38 collaboration \cite{na38} and the HELIOS-3 
collaboration \cite{cern} using a proton or sulfur projectile at 
200 AGeV and a tungsten or uranium target. A factor of 2 to 3 enhancement 
in the double ratio $( \phi / (\omega+\rho^{0} ))_{SU(W)} /( \phi / 
(\omega+\rho^{0}))_{pW}$ has been observed. Various theoretical attempts have
been made to understand this enhancement \cite{kosa,greiner,Koch}. In 
particular, an enhancement of phi meson yield may be a signature 
of the formation of a quark gluon plasma in the collisions \cite{shor}. 
However, the enhancement can also be explained 
in hadronic models if one takes into account the reduced phi meson mass 
in medium \cite{kosa}. In Ref. \cite{greiner}, the color rope fragmentation 
and hadronic rescattering have been introduced to understand 
the observed phi meson enhancement. 

Phi mesons have also been measured at AGS/BNL by the E-802 collaboration 
in central collisions between a 14.6 AGeV/c Si beam and a Au target 
\cite{AGS}. They are identified from the invariant mass spectrum 
of $K^{+} K^{-}$ pairs, and the measured phi meson mass and width 
are consistent with that in free space. The ratio of the 
phi meson yield to the $K^{-}$ yield has been found to be about 10\%, which 
can be understood if thermal and chemical equilibrium are assumed at
freeze out with a temperature of about 110 MeV \cite{bs95,fireball}. 
On the other hand, calculations by Baltz and Dover \cite{Dover}
based on the coalescence model, in which the phi meson is formed from 
the kaon and antikaon at freeze out, underestimate the data by a large factor.

Phi mesons from heavy-ion collisions at SIS/GSI energies are being 
studied by the FOPI collaboration \cite{FOPI} also through
the $K^{+} K^{-}$ invariant mass distribution. About 30 phi 
mesons have been identified in central collisions of Ni+Ni at 
1.93 AGeV. Based on these preliminary results it has been concluded 
that the phi meson yield is about 10\% of the $K^{-}$ yield, 
which is very similar to that observed at the AGS energies.
This is somewhat surprising since the SIS energies are below the phi meson 
production threshold in the nucleon-nucleon collision in free space, while 
the AGS energies are well above the threshold.     
 
In this paper, we shall carry out a detailed transport model study on 
phi meson production in heavy-ion collisions at SIS energies. There are 
mainly two reasons to work at these energies. First, 
particle production at subthreshold energies is sensitive to 
medium modifications of hadron properties \cite{mfrev,kkl,cass90,mos91}.  
Secondly, the reaction dynamics of heavy-ion collisions at SIS energies 
is relatively simple compared to that at higher energies as it involves 
mainly the nucleon, delta, and pion.  In a previous study 
of phi meson production at SIS energies \cite{kkonly}, only the contribution
from the reaction $K \bar K \to\phi$ has been included, and the 
$\phi / K^{-}$ ratio is predicted to be less than 1\%, which is an order 
of magnitude smaller than the experimental result from the FOPI 
collaboration. To explain this large discrepancy we include here 
phi meson production also from baryon-baryon and pion-baryon collisions. The 
cross sections for these processes are calculated using an 
one-boson-exchange model with parameters fitted to available experimental 
data.  Some results from this study have been reported 
earlier \cite{phibrief}.  We find that in heavy ion collisions the dominant 
contribution to phi meson production is from the $\pi N$ collision, 
and its inclusion in the transport mode brings the predicted $\phi/K^{-}$ 
ratio closer to the measured one by the FOPI collaboration.  Furthermore, 
we study the effect of medium modifications of kaon properties on phi meson 
production.  Kaplan and Nelson have pointed out 
that the kaon mass changes in nuclear matter \cite{N&K}. In chiral 
perturbation theory, the kaon mass is predicted to increase slightly while 
the antikaon mass decreases significantly as the density increases \cite{lee}. 
Overall, the total mass of a kaon-antikaon pair is lowered in dense matter. 
This affects appreciably the phi decay width $\Gamma_{\phi\rightarrow 
K{\bar K}}$, which is only about 3.7 MeV in free space,
since the phi meson mass is very close to twice the
kaon mass. A reduced kaon in-medium mass thus increases 
$\Gamma_{\phi\rightarrow K{\bar K}}$, so phi mesons are more likely to 
decay before freeze out, leading to a lower final phi meson yield.
This reduction is, however, compensated by effects from 
the dropping phi meson in-medium mass.

The paper is organized as follows: In section \ref{xsect} we calculate the 
elementary cross sections for phi meson production and absorption using 
an one-boson-exchange model. Section \ref{npik} describes the nucleon, 
pion and kaon dynamics in heavy ion collisions using the relativistic
transport model. Results on the phi meson yield and distributions 
from these collisions are given in section \ref{phiresult}. Section 
\ref{summary} is the summary and outlook.

\section{Elementary Cross Sections}
\label{xsect}
For heavy ion collisions at SIS energies, the most abundant particles 
are the nucleon, $\Delta$ resonance, and pion. 
Phi meson can thus be produced from 
reactions such as \nnnnphi, $\ndelnnphi$, $\deldelnnphi$, $\pinphin$,
and $\pidelnphi$. The production of phi meson from $\pi \pi$ scattering 
is expected to be unimportant due to its higher production threshold
than that for the other reactions, 
and is thus ignored. Although both kaon and antikaon are only scarcely
produced, phi meson production from the reaction $\kkphi$ turns out 
nonnegligible in the case of using in-medium kaon 
masses and will be included.

Unfortunately, the cross sections for these reactions are not
well-known empirically. There exist some data for the $\pi N$ channel near 
threshold \cite{pindata}, but the $N N$ channel has been studied 
only at high energies \cite{nndata}. Because of the scarcity of experimental 
data, an one-boson exchange model has been introduced in Ref. \cite{phibrief}
to extrapolate from the data at high energies to make predictions at 
energies near the production threshold, and also to calculate
the cross sections for reactions involving the delta
resonance as they cannot be measured experimentally. In that study, 
only the lowest order Born amplitude for phi meson production are 
considered, and this should be
a reasonable approximation to the full amplitude because phi meson production
has a very small probability in these reactions. 
In the following, we discuss this model and its predictions
for both phi production and absorption cross sections.

The interaction Lagrangians needed in the one-boson exchange model
are well-known, and they are given by
\begin{eqnarray}
{\cal L} _{\pi NN}& =& - {f_{\pi NN}\over m_\pi} {\bar \psi} \gamma ^5 \gamma
^\mu\vec\tau \psi\cdot \partial _\mu \vec\pi,\\
{\cal L} _{\rho NN}& =& -g_{\rho NN} {\bar \psi} \gamma ^\mu\vec\tau\psi
\cdot\vec\rho_\mu - {f_{\rho NN}\over 4m_N} {\bar \psi} \sigma ^{\mu\nu}
\vec\tau\psi\cdot (\partial _\mu\vec\rho_\nu -\partial _\nu\vec\rho_\mu),\\
{\cal L} _{\rho N \Delta}& =& i{f_{\rho N\Delta}\over m_\rho} {\bar \psi}
\gamma ^5 \gamma ^\mu\vec T \psi ^\nu\cdot(\partial_\mu\vec\rho_\nu
-\partial_\nu\vec\rho_\mu) + {\rm h.c.},\\
{\cal L_{\pirhophi}}& =& {f_{\pi \rho \phi} \over m_{\phi} }
\epsilon_{\mu \nu \alpha \beta} 
\partial^{\mu} \vec\phi^{\nu} \partial^{\alpha} \rho^{\beta}\cdot\vec\pi.
\end{eqnarray}
In the above, $\psi$ is the nucleon field with mass $m_N$; $\psi _\mu$ is 
the Rarita-Schwinger field for the spin-3/2 delta resonance with mass
$m_\Delta$; and $\pi$, $\rho_\mu$, $\phi_\mu$ are meson fields with masses
$m_\pi$, $m_\rho$, $m_\phi$, respectively.
The isospin matrices, Dirac $\gamma$-matrices, and Levi-Civita tensor
are denoted by $\vec\tau$, $\gamma^\mu$, and $\epsilon^{\mu\nu\alpha\beta}$,
respectively.
The coupling constants and cutoff parameters in ${\cal L}_{\pi NN}$,
${\cal L}_{\rho NN}$, and ${\cal L}_{\rho N\Delta}$
are taken from the Model II of Table B.1 in Ref. \cite{Mach}. 
From the measured width $\Gamma_{\phi\to\pi\rho}\approx 0.6$ MeV, 
the coupling constant $f_{\pirhophi}\approx 1.04$ is determined.

\subsection{Phi meson production from pion-baryon collisions}

\begin{figure}[p]
\begin{center}
\vfill
\mbox{\epsfxsize=14truecm\epsffile{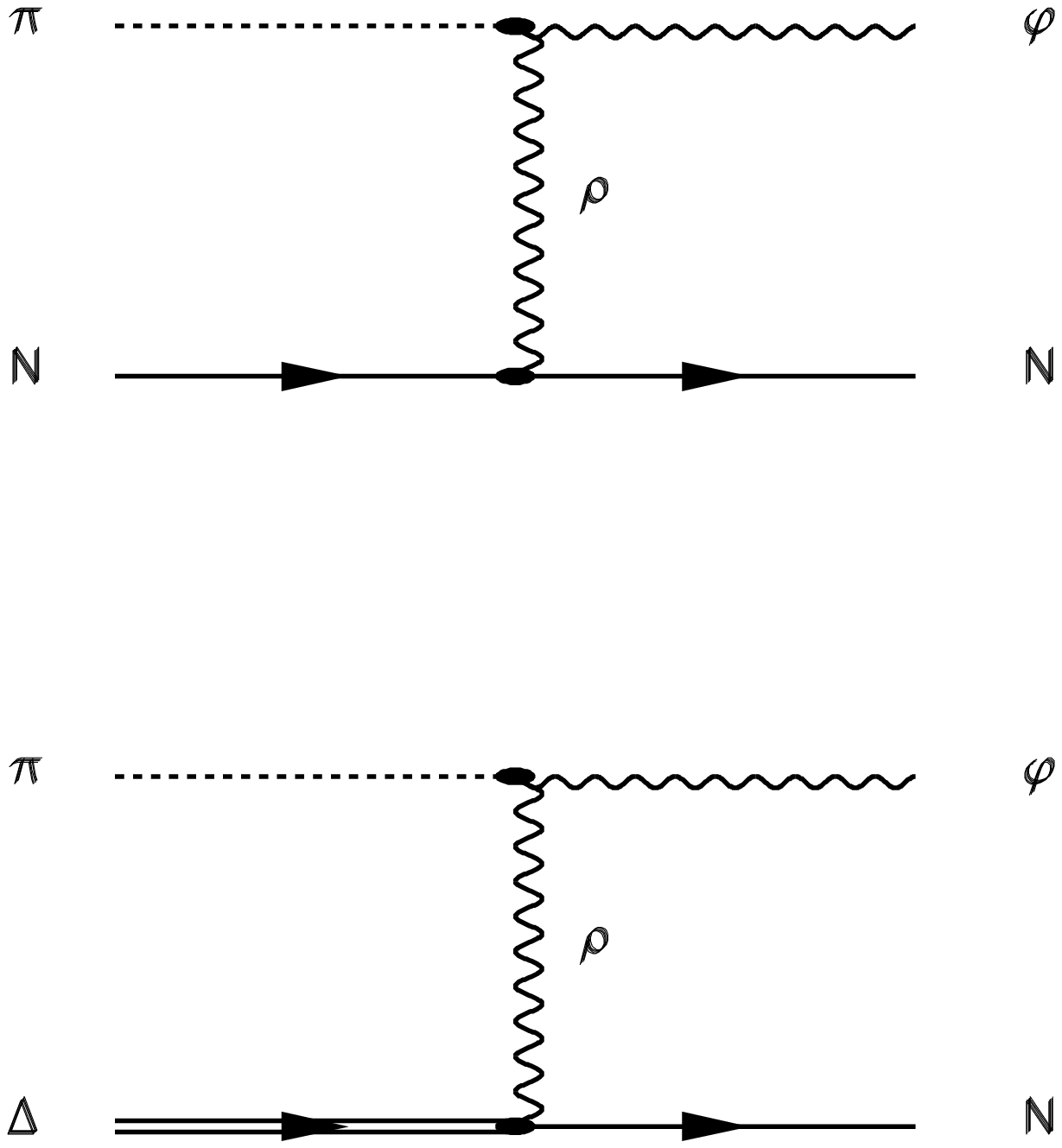}}
\caption{Feynman diagrams for $\pinphin$ and $\pidelnphi$. 
\label{FeynmanpiB} }  
\vfill
\end{center}
\end{figure}

The Feynman diagrams for the reactions $\pinphin \/$ and $\pidelnphi$ are 
shown in Fig. \ref{FeynmanpiB}.  A monopole
form factor is introduced at the $\pi\rho\phi$ vertex, i.e., 
\begin{equation}
 F(q_{\pi},q_{\rho}) = { {(\Lambda^\pi _{\pirhophi})^2 - m_{\pi}^{2}} \over 
            {(\Lambda^\pi _{\pirhophi})^2 + q_{\pi}^{2}} }.
\end{equation}
The cut-off parameter $\Lambda_{\pi\rho\phi}^\rho$ is adjusted to give a
reasonable fit to the experimental data from the reaction 
$\pi^{-}p \to n \phi$ \cite{pindata}, and a value of 1.2 GeV is obtained. 
The comparison between calculated results and the experimental data 
as well as the parameterization by Sibirtsev \cite{sib}
can be found in Ref. \cite{phibrief}.

\begin{figure}[p]
\begin{center}
\vfill
\mbox{\epsfxsize=14truecm\epsffile{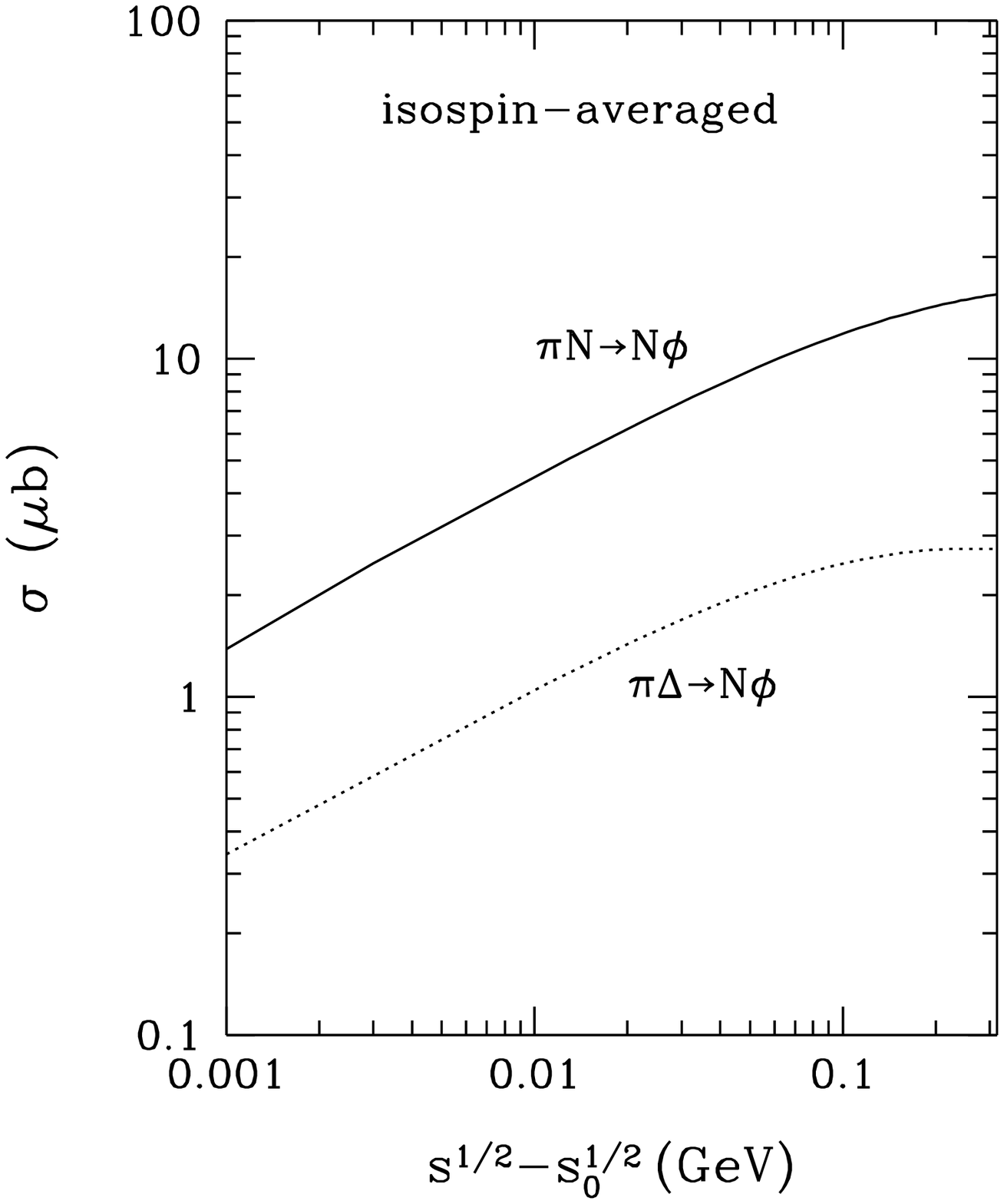}}
\caption{Isospin-averaged cross sections for 
$\pinphin$ (solid curve) and $\pidelnphi$ (dotted curve).
\label{isopb} }
\vfill
\end{center}
\end{figure}

Since the isospin degree of freedom is usually not explicitly treated
in transport model, we have thus calculated the isospin-averaged cross 
sections for $\pinphin \/$ and $\pidelnphi$.
The results are shown in Fig. \ref{isopb}. 
Near threshold the cross section for $\pinphin$ is seen to be much larger 
than the cross section for $\pidelnphi$ as a result of the strong tensor
coupling at the $\pi NN$ vertex.

\begin{figure}[p]
\begin{center}
\vfill
\mbox{\epsfxsize=14truecm\epsffile{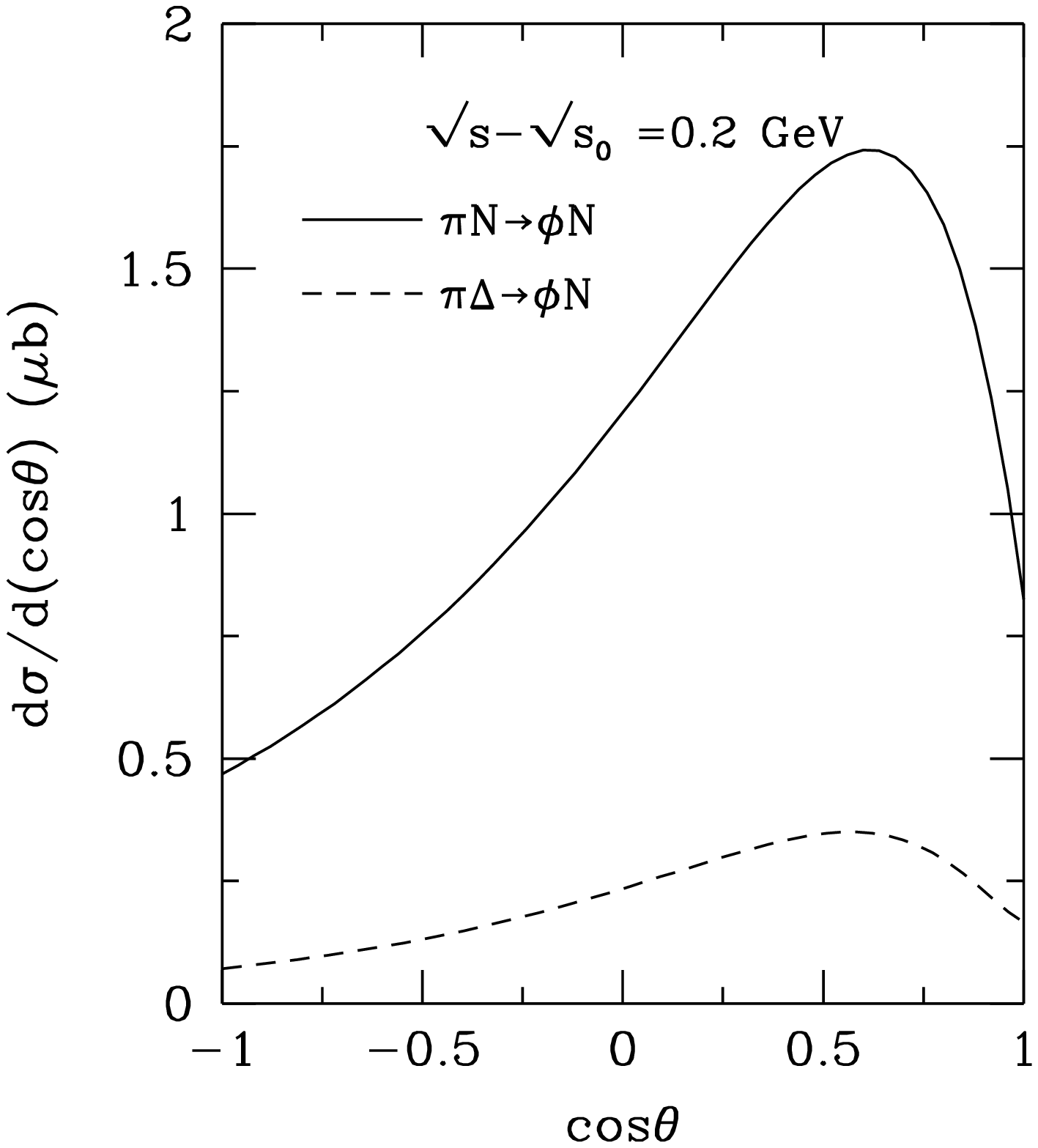}}
\caption{Differential cross sections for $\pinphin$ (solid curve) 
and $\pidelnphi$ (dashed curve) at 0.2 GeV above the threshold. 
\label{diffXpiB} } 
\vfill
\end{center}
\end{figure}

To determine the momentum of produced phi meson in the
transport model, we also need the differential cross sections
for these reactions. Since only two particles are in the final state,
the magnitude of the
phi momentum is fixed by kinematics. The differential cross
section $d\sigma /dcos\theta$ for these reactions at an 
energy of 0.2 GeV above the threshold is shown in Fig. \ref{diffXpiB}. 

\subsection{phi meson production from baryon-baryon collisions}
\label{pbb}	
\begin{figure}[p]
\begin{center}
\vfill
\mbox{\epsfxsize=14truecm\epsffile{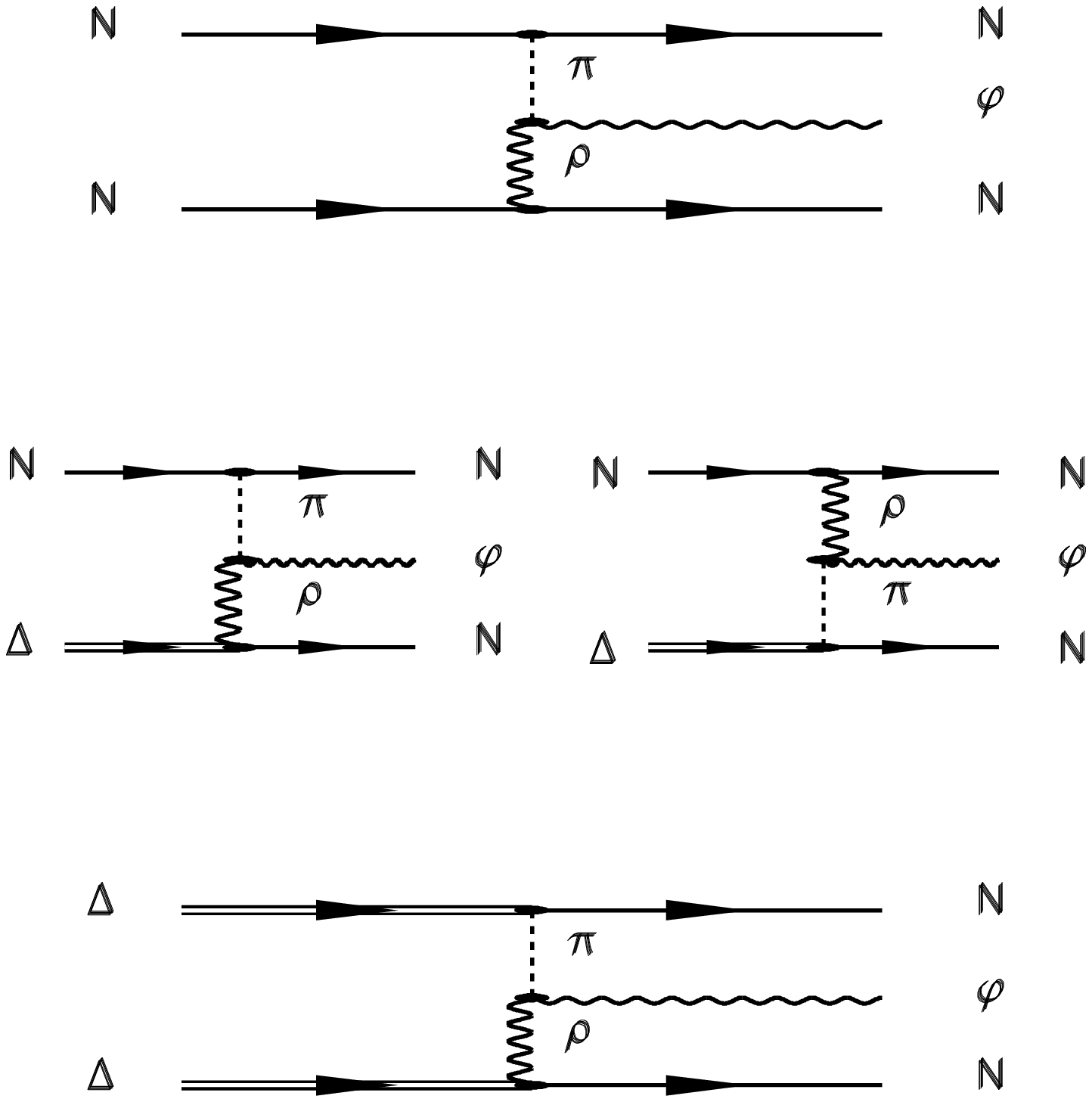}}
\caption{Feynman diagrams for $\nnnnphi$, $\ndelnnphi$, and $\deldelnnphi$. 
\label{bb} }  
\vfill
\end{center}
\end{figure}

The Feynman diagrams for the reactions 
$\nnnnphi$, $\ndelnnphi$, and $\deldelnnphi$ are shown in Fig. \ref{bb}.
Exchange diagrams are also included as the 
two final nucleons are identical. 
There is some ambiguity in treating the $\pirhophi$ vertex
in Fig. \ref{bb}, which involves two
virtual particles. Such a vertex does not exist in the Bonn potential 
model, and there is also no theoretically well-defined criteria for 
prescribing its form. The following form has been used in Ref. \cite{phibrief},
\begin{equation}
 F(q_{\pi},q_{\rho}) = 
        \left( { {(\Lambda^\pi _{\pirhophi})^2 - m_{\pi}^{2}} \over 
            {(\Lambda^\pi _{\pirhophi})^2 + q_{\pi}^{2}} } \right)
        \left( {{(\Lambda^\rho _{\pirhophi})^2 - m_{\rho}^{2}} \over 
            {(\Lambda^\rho _{\pirhophi})^2 + q_{\rho}^{2}}} \right), 
\end{equation}
where $q_{\rho}$ and $q_{\pi}$ are four momenta of the virtual
$\rho$ and $\pi$ meson, respectively. The cut-off parameter 
$\Lambda^{\rho}_{\pirhophi}$ is the same one introduced earlier in 
calculating the cross section for $\pinphin$.  
The other cut-off parameter $\Lambda^{\pi}_{\pirhophi}$ 
is introduced to take into account the virtuality of the pion.
The cross section for $pp \arr p p \phi \/$ has been measured at several 
energies \cite{nndata}, and the one at the lowest energy has been used
in Ref. \cite{phibrief} 
to fix $\Lambda^{\pi}_{\pirhophi}$, and a value of 0.95 GeV has been obtained.
Comparisons of calculated results with both experimental data 
and the results by Sibirtsev based on an one-pion exchange model \cite{sib}
have been given in 
Ref. \cite{phibrief}. We note that both $\Lambda^\rho_{\pi\rho\phi}$ 
and $\Lambda^{\pi}_{\pirhophi}$  are in the order of 1 GeV, 
similar to typical values for cut-off parameters in both the Bonn potential
model and other hadronic models. 

For the reactions $\ndelnnphi$ and $\deldelnnphi$, the exchanged pion 
can be on shell, so the cross sections are singular in certain kinematical
region.  This singularity can be regulated by the Peierls method
\cite{peierls}, in which the energy of the delta resonance is taken
to be complex to account for its finite lifetime. As a result, 
the four-momentum of the exchanged pion acquires an imaginary part, 
which thus renders the pion propagator finite at the pion pole. 

\begin{figure}[p]
\begin{center}
\vfill
\mbox{\epsfxsize=14truecm\epsffile{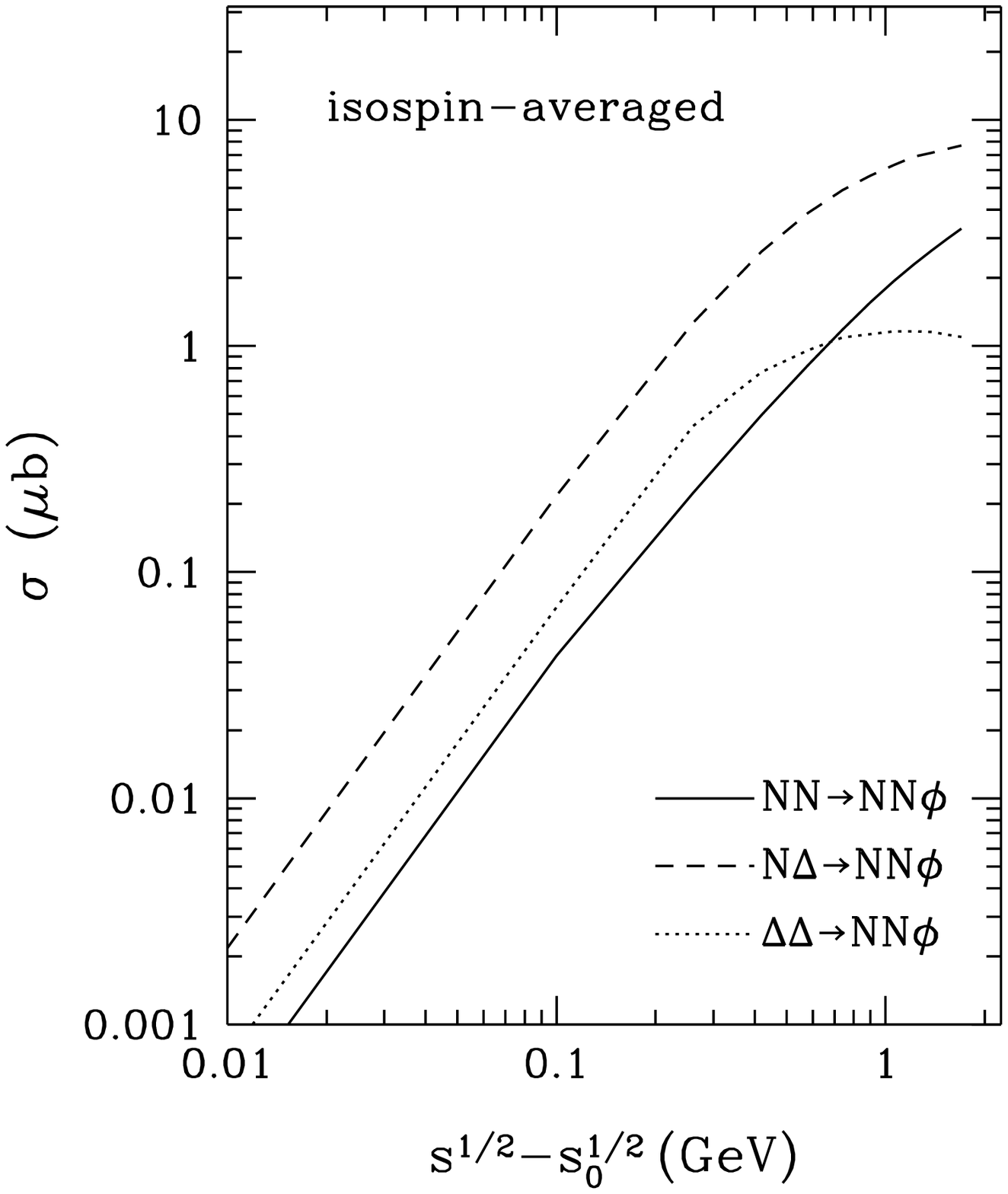}}
\caption{Isospin-averaged cross sections for $\nnnnphi$ (solid curve),
$\ndelnnphi$ (dashed curve), and $\deldelnnphi$ (dotted curve).
\label{isobb} }
\vfill
\end{center}
\end{figure}

The isospin averaged cross sections for phi meson production from 
baryon-baryon collisions are shown in Fig. \ref{isobb}. 
There are significant differences between the cross sections for
the reactions $\nnnnphi$, $\ndelnnphi$, and $\deldelnnphi$. 
Near threshold, 
all are much smaller than the cross section for phi meson production from
pion-nucleon collisions.

\begin{figure}[p]
\begin{center}
\vfill
\mbox{\epsfxsize=14truecm\epsffile{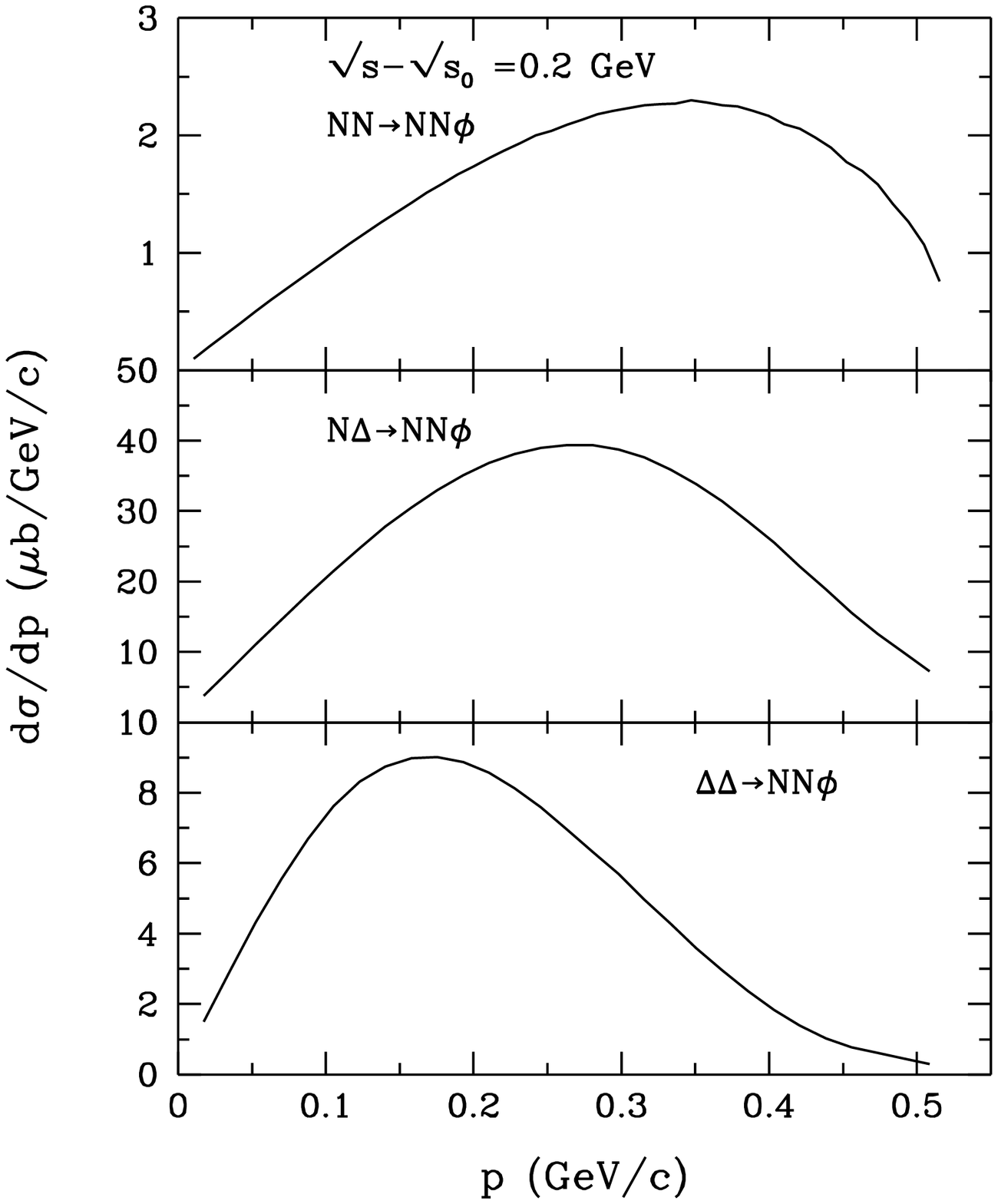}}
\caption{Phi meson momentum spectrum from the reactions $\nnnnphi$,
$\ndelnnphi$, and $\deldelnnphi$ at 0.2 GeV above the threshold. 
\label{pspectrum} }
\vfill
\end{center}
\end{figure}

Since there are three particles in the final state in baryon-baryon 
collisions, the momentum of produced phi meson is given by 
the triple differential cross sections 
$d^3\sigma /dp_xdp_ydp_z$ or double differential cross
section $d^2\sigma /p^2dpd\Omega$ if there is an azimuthal symmetry.
To simplify the numerical
procedure, we introduce the assumption that the phi meson momentum is
isotropic in baryon-baryon center-of-mass frame. We then
need only the momentum spectrum $d\sigma /dp$, which is shown 
in Fig. \ref{pspectrum} for these reactions again at an energy of 0.2 GeV
above the threshold.  

\subsection{Phi meson production from kaon-antikaon annihilation}

The total cross section for $\kkphi$ is assumed to have
a Breit-Wigner form as in Ref. \cite{kkonly}, i.e., 
\begin{equation}\label{kkp}
   \sigma ( K \bar K \arr \phi) = {3 \pi \over k^2} 
    { ( m_{\phi} \Gamma_{\phi} )^2 \over 
        (M^2-m_{\phi}^2)^2+(m_{\phi}\Gamma_{\phi})^2 },
\end{equation}
\noindent where $\Gamma_{\phi}$ is the phi meson decay width to 
$K\bar K$ and is given by
\begin{equation} 
  \Gamma_{\phi} = { g^{2}_{\phi K {\bar K}} \over 4 \pi }
       { (m_\phi^{2}-4m_{K}^{2})^{3/2}  \over 6 m_\phi^2},
 \label{phiwidth}
\end{equation}
with the coupling constant ${g^{2}_{\phi K {\bar K}}/ 4 \pi } \approx 
1.69$ determined from the empirical width of 3.7 MeV. In using Eq. 
(\ref{kkp}), we have thus neglected the small effect due to the decay of
phi meson into other channels.

\subsection{Phi meson scattering by nucleons}

\begin{figure}[p]
\begin{center}
\vfill
\mbox{\epsfxsize=14truecm\epsffile{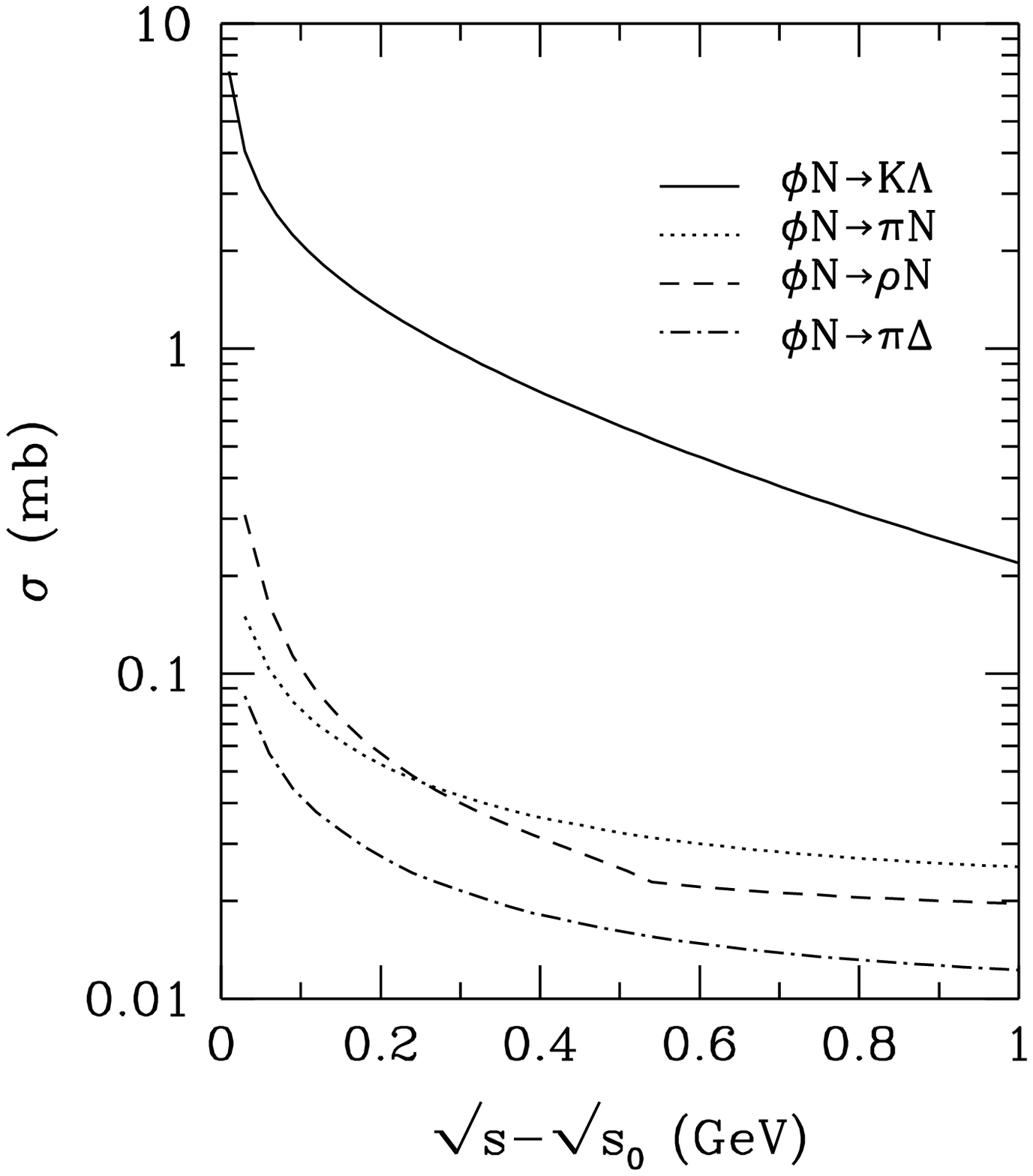}}
\caption{Cross sections for $\phinlambdak$ (solid curve),
$\phinpin$(dotted curve), $\phinnrho$ (dashed curve), and 
$\phinpiDel$ (dash-dotted curve). 
\label{phiabs}}
\vfill
\end{center}
\end{figure}

Once a phi meson is produced, there are three kinds of reactions it can 
take part in: decay, inelastic and elastic scattering by nucleons. The 
dominant decay mode of phi meson is $\phi \to K {\bar K}$, with a decay width  
given by Eq. (\ref{phiwidth}). Here we only consider phi meson absorption 
due to $\phi N$ collisions as that due to $\phi\pi$ and $\phi \Delta$ 
scattering is expected to be insignificant because of the low 
pion and delta densities compared with that of nucleons (see Fig. 
\ref{cden}). We have included the reactions $\phinlambdak$, $\phinnrho$, 
$\phinpin$, and $\phinpiDel$. The cross section for the first reaction is 
related via detailed balance to that of $\Lambda K\to\phi N$, which has
already been calculated in Ref. \cite{kosa} based on a kaon exchange model
with parameters taken from Ref. \cite{holzen}. The reaction
$\phi N\rightarrow K\Sigma$ is neglected since the 
$N\Sigma K$ coupling constant ($g_{N\Sigma K}^2/4\pi \sim 0.6$) is much 
smaller than that of $N\Lambda K$ ($g_{N\Lambda K}^2/4\pi \sim 16$)
\cite{holzen}. 
The cross section for the other three reactions can be obtained 
from those for $\rho N\to\phi N$, $\pi N\to\phi N$, and $\pi\Delta\to\phi N$ 
using detailed balance relations. We have already determined in the
above the cross sections for phi meson production from pion-baryon collisions.
For the reaction $\rho N\to\phi N$, we simply interchange the pion and
rho meson in the first diagram of Fig. \ref{FeynmanpiB}.  
The cross sections for these phi meson absorption processes are 
summarized in Fig. \ref{phiabs}. We see that $\phinlambdak$ 
is the dominant one as already pointed out in Refs. 
\cite{shor,kosa}, because this is the only one 
among the four reactions that is not suppressed by the OZI rule . 

The phi-nucleon elastic scattering is also included. Experimentally, the 
cross section for this process can be extracted from data on phi meson 
photoproduction using the vector meson dominance model \cite{joo}. 
In the energy range $3.5$ GeV$<p_{\rm lab}<5.8$ GeV, a value of 
\mbox{0.56 mb} has been obtained. Here we make the rough approximation 
that the cross section is a constant of 0.56 mb in the energy range 
considered.  We note that a slightly larger value has been obtained in Ref. 
\cite{barger}. The phi-nucleon scattering does not change the phi meson
yield, but affects its momentum distribution. Since this cross section 
is rather small, the effect is not appreciable as shown below.

\subsection{Discussions}

So far we have calculated exclusive phi meson production cross sections
with no pions in the final state. In principle, phi meson can also be
produced in association with one or more pions, and one needs for
heavy-ion collisions inclusive cross sections such 
as $NN\rightarrow \phi X$. Since the beam energies we consider
in this work are below the phi meson production threshold
in nucleon-nucleon collision, we expect that exclusive
reactions such as $NN\rightarrow NN\phi$, which have lower 
thresholds, are more important than the ones with pions
in the final state, which require additional energies
and are thus suppressed. 

We would like to point out that in spite of the success 
of the one-boson-exchange model at low energies, this model has not been
carefully verified at energies above the phi meson production threshold. 
It is often believed that the quark degree of freedom becomes relevant 
at a kinetic energy of 1-2 GeV, so it is not known if our procedure of using 
the data point at $p_{\rm lab}=10$ GeV to fix the cross section for 
$\ppppphi$ near the threshold is reliable. Nevertheless, this model
has been used quite often in calculating the cross sections for
pion production \cite{pion}, 
eta production \cite{mosel}, kaon production \cite{laget,liko}, and 
baryon resonance excitations \cite{huber} from nucleon-nucleon collisions
in this energy region. Measurements of $pp \to pp\phi$ at low 
energies are being carried out at SATURNE \cite{saturne}, and these data
will be very useful in testing the validity of our model.
With this caveat in mind, we proceed to our 
transport model study using the above elementary cross sections.

\section{The nucleon, pion and kaon dynamics}
\label{npik}

\subsection{The relativistic transport model}

For the description of heavy-ion collisions at SIS energies, we use the 
relativistic transport model (RVUU) developed in Ref. \cite{RVUU}
from the non-linear $\sigma-\omega$ model \cite{qhd}
and extensively used in previous studies. At these 
energies, the colliding system consists mainly of nucleons, delta 
resonances, and pions.  While medium effects on pions are neglected 
as in most transport models, nucleons and delta resonances propagate 
in a common mean-field potential according to Hamilton equations
of motion, i.e., 
\begin{eqnarray}\label{EOMB}
{d{\bf x}\over dt}=~{{\bf p}^*\over E^*} \qquad\qquad
{d{\bf p}\over dt}=~-\nabla_{\bf x} (E^*+(g_\omega /m_\omega )^2\rho _B),
\end{eqnarray}
where $E^*=(m^{*2}+{\bf p}^{*2})^{1/2}$. The effective mass and kinetic 
momentum of a baryon are given, respectively, by
$m^*=m-g_\sigma\langle\sigma\rangle$ and
${\bf p}^*={\bf p}-g_\omega\langle{\bbox\omega}\rangle$, where
the expectation values $\langle\sigma\rangle$ and 
$\langle{\bbox \omega}\rangle$ 
are related to the attractive scalar potential
and the vector current of a baryon in nuclear matter. We use in this 
work the parameter set that corresponds to the soft equation of
state \cite{liko94} with a compressibility $K=200$ MeV
and an effective nucleon mass $m_N^*=0.83~m_N$ at normal nuclear density.
These particles also undergo stochastic two-body 
collisions, including both elastic ($NN\rightarrow NN$, 
$N\Delta\rightarrow N\Delta$, 
$\Delta\Delta \rightarrow \Delta\Delta$) and inelastic ($NN\leftrightarrow 
N\Delta$, $\Delta\leftrightarrow N\pi$) scattering.
The standard Cugnon parameterization 
\cite{bert} and proper detailed-balance prescription \cite{dan} are 
used for describing these reactions.

Besides pions, other particles are also produced in heavy-ion collisions. 
However, their production probabilities are very small at these energies,
and they can thus be treated perturbatively, in the sense that their 
production and interactions do not affect the dynamics of nucleons, delta 
resonances, and pions.  This is carried out using the perturbative test 
particle method of Ref. \cite{fang93,Fang}. In the case of kaons, they are
produced from pion-baryon and baryon-baryon collisions whenever the energy 
is above the threshold. Then, a probability factor given by the kaon 
production cross section to the total pion-baryon or baryon-baryon cross 
section is assigned to the produced kaon. The motion and collision of 
the kaon are then followed. However, only the kaon momentum is changed 
after a kaon-nucleon collision.  This method allows us to include easily 
the medium effects on the kaon as shown in Ref. \cite{fang93}. We use 
in this study the kaon in-medium mass obtained from the mean-field 
approximation to chiral Lagrangian 
\cite{N&K}.  As in Ref. \cite{Kflow}, the kaon mass $m_K^*$
at nuclear matter density $\rho$ is given by  
\begin{equation}\label{kmass}
m_K^*=m_K\left[1-\frac{\Sigma_{KN}}{f^2m_K^2}\rho_s+(\frac{3}{8}\frac{\rho}
{f^2})^2\frac{1}{m_K^2}\right]^{1/2}+\frac{3}{8}\frac{\rho}{f^2}.
\end{equation}
In the above, $m_K$ is the kaon mass in free space; $f\approx 93$ MeV
is the pion decay constant; and $\rho_s$ is the nuclear scalar density. The  
kaon-nucleon sigma term is denoted by $\Sigma_{KN}$ and is taken to be
about 350 MeV. At normal nuclear density, the kaon mass is then increased
by about 10 MeV. For antikaon, its vector interaction is opposite to that
of kaon, so the last term in Eq. (\ref{kmass}) has instead a negative sign.
The antikaon mass at normal nuclear density is reduced by about 100 
MeV.  Corrections to the mean-field results may not be negligible
\cite{lee,waas} and are still under debate. Since there are other
uncertainties in the treatment of kaons such as their production cross
sections from baryon-baryon collisions, we will not address the sensitivity
of our results on these corrections.  

Our treatment of kaon follows essentially Refs. \cite{Fang,Kflow,lieos}, 
where the baryon-baryon production channels $B B \to N\Lambda K$ and $B B 
\to N\Sigma K$ are included with their cross sections given by 
the Randrup and Ko parameterization \cite{ran}.  We have also
included the pion-baryon channels $\pi N \to\Lambda K$ and $\pi N 
\to\Sigma K$. For these cross sections, the parameterizations due to 
Cugnon \cite{Cugnon} are used.  Kaon production from $\pi\Delta$ collisions
is neglected as its cross section has been shown to be much smaller 
than that for $\pi N$ collisions \cite{fae}.

The production of antikaon in heavy-ion collisions at SIS
energies was first studied in Ref. \cite{akaon} using the
relativistic transport model and the Zwermann-Sch\"urmann \cite{zwer}
parameterization for the antikaon production cross section
from the nucleon-nucleon collision. However, it has been recently
found that this parameterization overestimates the cross sections
near the threshold by about one order of magnitude \cite{sibirtsev,cass}, 
and contributions from pion-hyperon interactions \cite{ko83}, 
which were neglected in Ref. \cite{akaon},
are needed to account for the measured $K^-$ yield. 
In this work, we have thus modified our earlier treatment
of antikaon production by using instead the parameterizations proposed
in Ref. \cite{sibirtsev} for antikaon production cross sections
in both pion-baryon and baryon-baryon collisions. We have
also included contributions from the pion-hyperon collisions. 

\subsection{The pion and nucleon yield and distributions}

\begin{figure}[p]
\begin{center}
\vfill
\mbox{\epsfxsize=14truecm\epsffile{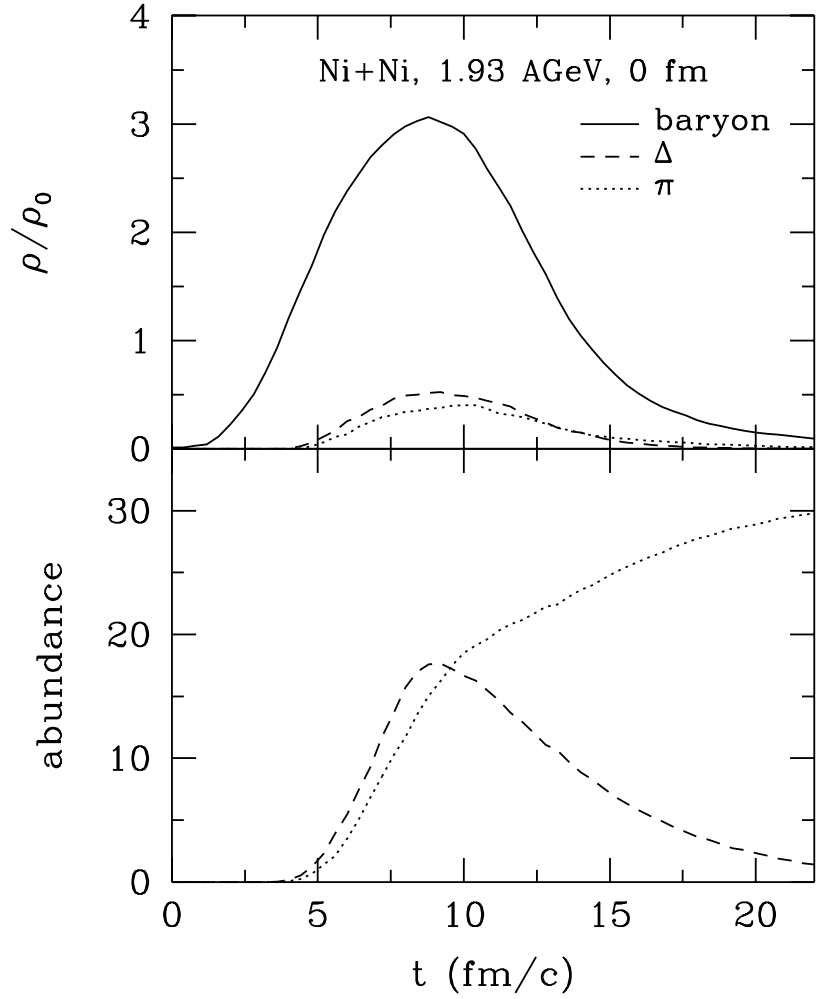}}
\caption{Upper window: the time evolution of central densities
of baryons (solid curve), deltas (dashed curve), and pions (dotted curve)
in Ni+Ni collisions at 1.93 AGeV and impact parameter 0 fm.
Lower window: the time evolution of the delta (dashed curve) and pion 
(dotted curve) abundance. 
\label{cden}}
\vfill
\end{center}
\end{figure}

In the upper window of Fig. \ref{cden}, we show the time evolution 
of central baryon and pion densities in Ni+Ni collisions 
at an energy of 1.93 AGeV and zero impact parameter. It is 
seen that the maximum baryon density reached in the 
central region is around three times normal nuclear 
density, while the maximum delta and pion densities are about 1/6
of the maximum nucleon density. The abundance of pions and deltas 
is given in the lower window of Fig. \ref{cden},  which shows 
that the maximum number of delta is around 18. Since the 
number of pions reaches around 30 at freeze out, the 
pion/nucleon ratio is thus about 0.25, which is in agreement 
with the experimental data \cite{piond} as well 
as other transport model results \cite{Wolf}. 

\begin{figure}[p]
\begin{center}
\vfill
\mbox{\epsfxsize=14truecm\epsffile{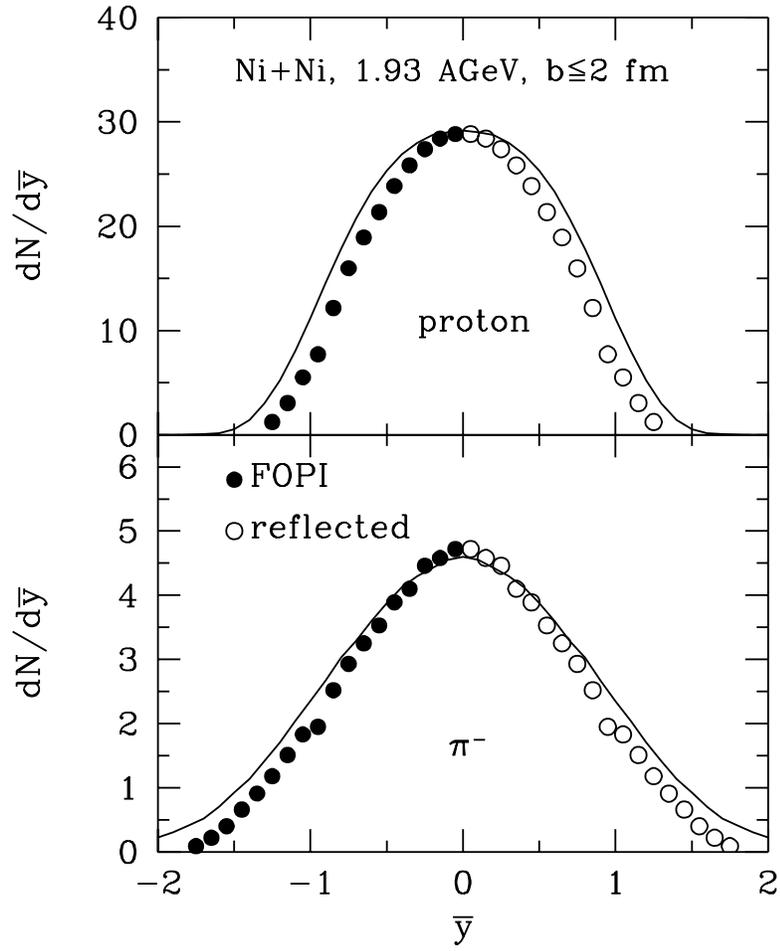}}
\caption{The rapidity distribution of nucleons
and pions in Ni+Ni collisions at 1.93 AGeV
and impact parameter ${\rm b}\le 2$ fm. 
\label{npiy} }
\vfill
\end{center}
\end{figure}

Our results for proton and $\pi^-$
rapidity distributions are compared with the experimental
data from the FOPI collaboration \cite{FOPI} in Fig. \ref{npiy}, where 
${\bar y}=y/y_{\rm projectile}$ is the normalized center-of-mass rapidity. 
The theoretical results are obtained for the impact parameter range $b\le 2$ 
fm in order to compare appropriately with the data which have a centrality 
selection of 100 mb. Also, only protons that have a local density less
than 0.02 fm$^{-3}$ at freeze out are included. Otherwise, they are
considered as bounded in the deuteron and other light fragments. The 
agreement between theoretical results and data, especially in 
mid-rapidity region, is quite good, so we feel that
our transport model describes properly the dynamics of nucleons, deltas, 
and pions. The proton rapidity distribution shows a peak at
${\bar y}=0$, thus indicating that considerable amount of stopping is
achieved in the collisions. 
 
\begin{figure}[p]
\begin{center}
\vfill
\mbox{\epsfxsize=14truecm\epsffile{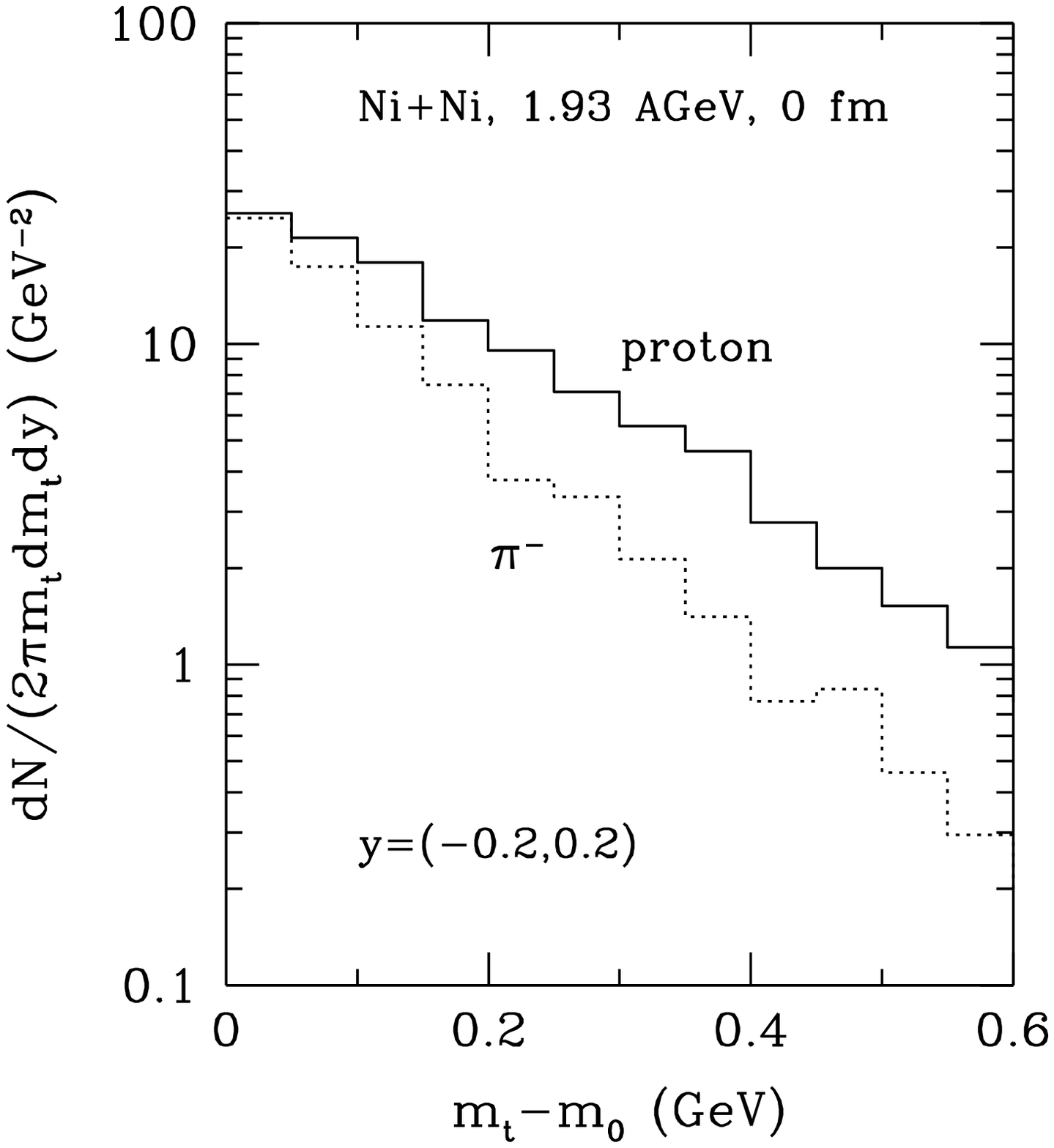}}
\caption{Same as Fig. \protect\ref{npiy} for the transverse mass
distribution. 
\label{npimt} } 
\vfill
\end{center}
\end{figure}

In Fig. \ref{npimt}, the nucleon and pion transverse mass 
distributions at mid-rapidity region $(-0.2<y<0.2)$ are shown.  
We find that the inverse slope parameter for nucleons 
is about 180 MeV, while that of pions is about 120 MeV. The difference is due 
to radial flow, which has a larger effect on
nucleons than pions.

\subsection{The kaon yield and distribution}
\label{kaon}
\begin{figure}[p]
\begin{center}
\vfill
\mbox{\epsfxsize=14truecm
\epsffile{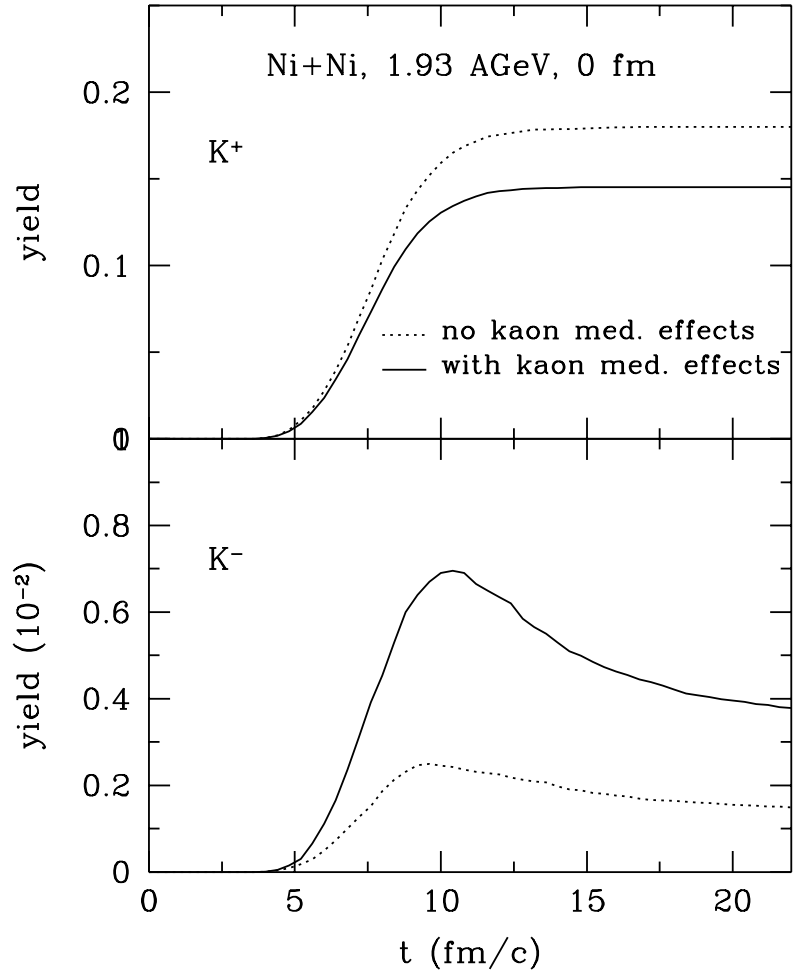}}
\caption{The abundance of $K^+$ and $K^-$ with (solid curve) and without
(dotted curve) medium effects in the same
reaction as in Fig. \protect\ref{cden}.
\label{kabnd}} 
\vfill
\end{center}
\end{figure}

The abundance of kaons and antikaons is shown in Fig. \ref{kabnd}. 
Without kaon medium effects the $K^+$ number reaches 
about 0.18, while that of $K^-$ reaches about $1.4 \times 10^{-3}$. 
Including kaon medium effects, the kaon yield is reduced by about 
20\% to about 0.15,  while the $K^-$ yield increases by about 
a factor of 3 to about $4 \times 10^{-3}$.
In our model, $K^+$ feels a weak repulsive potential in nuclear
matter, which increases its production threshold and thus reduces
the yield slightly. On the other hand, $K^-$ has a
strong attractive potential, so its production threshold is reduced
and a much larger yield is obtained.

\begin{figure}[p]
\begin{center}
\vfill
\mbox{\epsfxsize=14truecm\epsffile{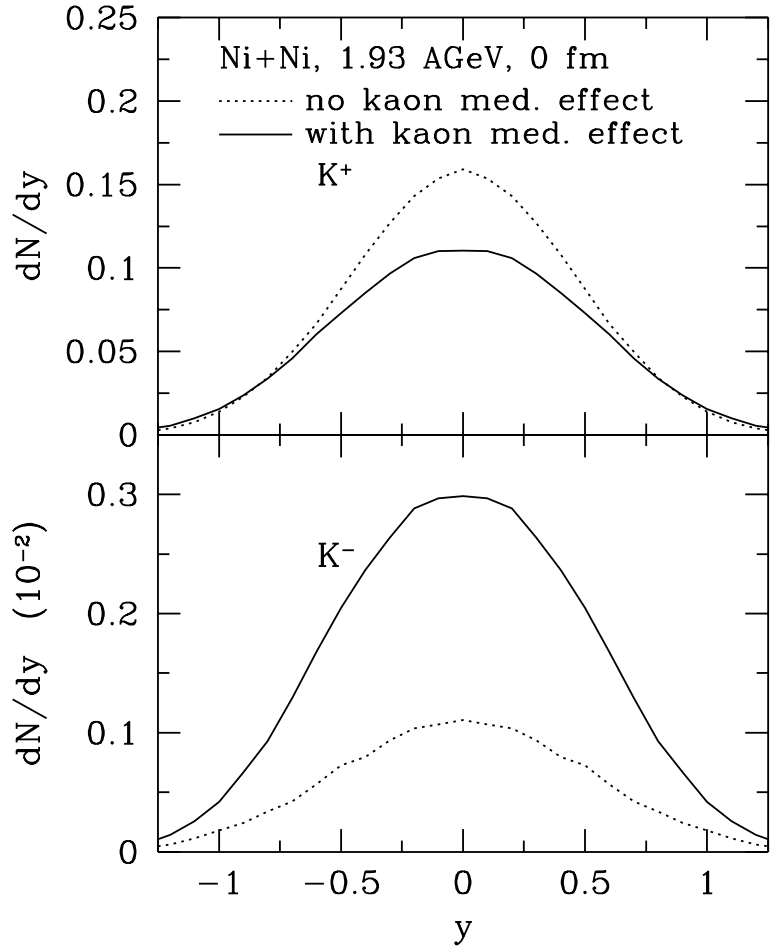}}
\caption{The rapidity distribution of $K^+$ and $K^-$ in 
central Ni+Ni collisions at 1.93 AGeV with (solid curve) and without 
kaon (dotted curve) medium effects. 
\label{ky} }
\vfill
\end{center}
\end{figure}

\begin{figure}[p]
\begin{center}
\vfill
\mbox{\epsfxsize=14truecm\epsffile{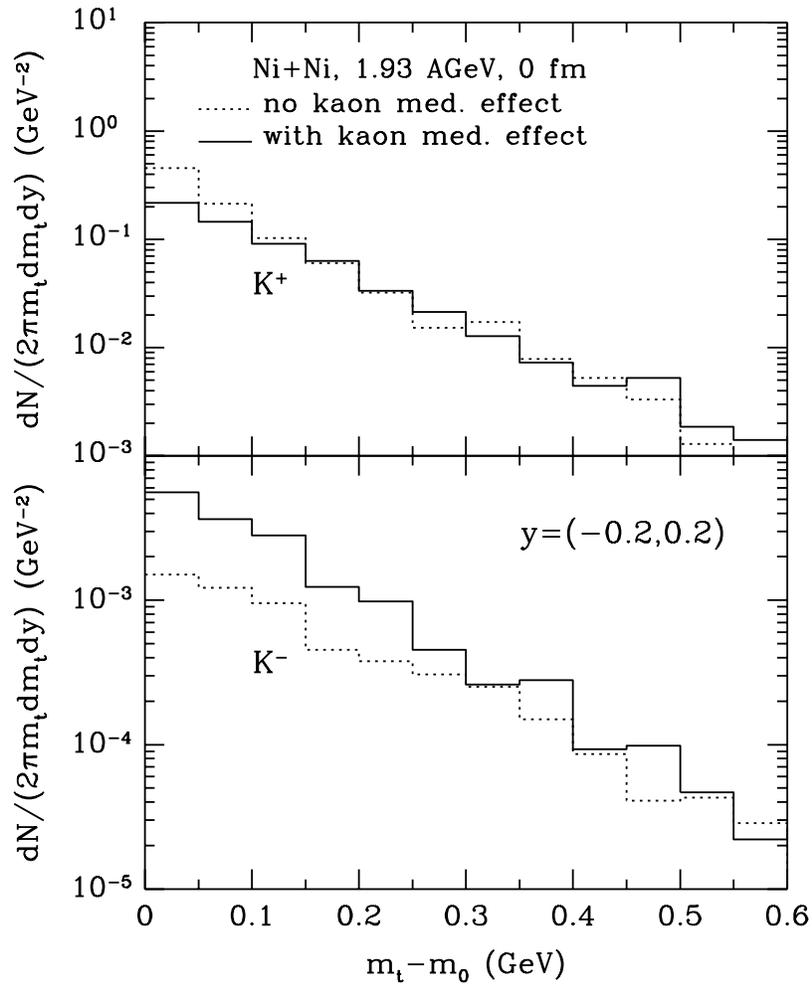}}
\caption{Same as Fig. \protect\ref{ky} for the transverse mass
distribution of $K^+$ and $K^-$ with (solid curve) and without (dotted
curve) medium effects.	
\label{kmt} }
\vfill
\end{center}
\end{figure}

The kaon and antikaon rapidity distributions for central Ni+Ni 
collisions are shown in Fig. \ref{ky} for both cases with 
and without kaon medium effects.  As in the case of pions 
and nucleons, the rapidity distributions peak at $y=0$. 
In Fig. \ref{kmt}, the kaon and antikaon transverse mass 
distributions in mid-rapidity region $(-0.2 < y < 0.2)$ are 
shown. For kaon, the inverse slope parameter is found to be about 
100 MeV without kaon medium effects, and increases to about 
120 MeV after kaon medium effects are included. The kaon momentum thus 
increases as a result of propagation in the repulsive mean-field potential. 
For antikaon, an inverse 
slope parameter of about 110 MeV is found in the case without 
kaon medium effects. This is somewhat larger than that of the 
kaon transverse mass spectra, mainly due to the absorption of 
low-momentum antikaons by nucleons because of a larger absorption 
cross section \cite{akaon}. Including kaon medium effects, 
the inverse slope parameter decreases to about 100 MeV. The propagation 
in the mean-field potential thus reduces the antikaon momentum 
as a result of the attractive potential. Therefore, from the ratio 
of kaon and antikaon transverse mass spectra it is possible to learn
about the kaon medium effects as pointed out in Ref. \cite{koch}.  

\section{Phi meson production from heavy ion collisions}
\label{phiresult}

\subsection{Transport model for the phi meson}

Phi mesons are also treated in our model by the perturbative test particle
method \cite{fang93}.  Whenever a pion-baryon or a baryon-baryon collision
is above the threshold for phi meson production, a phi meson is produced and
is assigned a probability factor that is given by the phi meson production
cross section to the total pion-baryon or baryon-baryon cross section. 
The motion of the phi meson, including its collisions with other nucleons,
is then followed. As in the case of kaon, effects of these collisions 
are included for phi mesons but not nucleons. The numerical simulation 
of heavy ion collisions is stopped when the number of collisions is 
small. The number 
of phi mesons at this time is taken to be the yield accessible 
to measurement by identifying $K \bar K$ pairs. As shown in 
Refs. \cite{Fang,akaon}, $K N$ and ${\bar K} N$ scattering are 
significant before freeze out, so the $K \bar K$ pair coming 
from the phi meson decay inside the fireball loose their
correlation before they get out and cannot be identified as a phi 
meson from the $K \bar K$ invariant mass spectrum.

In $\pi N$ or $\pi \Delta$ channels phi meson is produced in a 
two-body final state, so the magnitude of its momentum 
is fixed by kinematics. The polar angle of the phi meson momentum
depends on the differential cross section and is thus determined
by a probability function given by the ratio of the differential 
cross section to the total cross section.  For the azimuthal angle of the 
momentum, it has a uniform distribution between zero and $2\pi$.  

The baryon-baryon channels are more complex as the phi meson is produced 
in a three-body final state. In this case, the final momentum of the 
phi meson follows a multivariable probability function given by the 
ratio of the differential cross section to the total cross section.
To simplify the numerical procedure,  
we assume that the phi meson is produced isotropically in the 
baryon-baryon center of mass frame as discussed in Section \ref{pbb}. This 
assumption is often used in 
transport model study when no data exist for the differential cross section. 
The magnitude of phi meson momentum is then determined by 
the momentum spectrum such as that shown in Fig. \ref{pspectrum}

When including medium effects, we only change the available phase 
space due to the modification of hadron masses.
Modifications on coupling constants, cut-off parameters, and the functional
form of cross sections are neglected in this study. In the case of a 
reaction $A + B \to C + D$, the functional form of the cross section as a 
function of $E(=\sqrt{s^*}-\sqrt{s_0^*})$ is assumed to be the same as in 
free space, where the modified center of mass $\sqrt{s^{*}}$ energy is 
given by $\sqrt{s^{*}} = \sqrt{ p_{A}^{*2}+m_{A}^{*2} } + 
\sqrt{p_{B}^{*2}+m_{B}^{*2} }$, while the modified production threshold is 
$\sqrt{s^{*}_{o}}=m_{C}^{*2}+m_{D}^{*2}$,
with $p^*$ and $m^*$ being the momentum and mass of 
a particle in nuclear medium. 

In our study medium modifications of pion properties are
neglected, while those on nucleons are always 
included. For the treatment of kaon, antikaon and phi meson
production, we will consider three scenarios. In the first
scenario, we neglect medium effects on both kaons and phi mesons.
It is already known from previous studies of antikaon production
\cite{akaon,cass} that the neglect of kaon medium effects
underestimates the antikaon yield. In the second scenario,
we include medium effects on kaons but neglect those on phi mesons. The 
kaon medium effects are obtained from the chiral perturbation
calculation as discussed earlier. Inclusion of kaon medium effects improves
the agreement of the antikaon yield with the experimental data
as discussed in Section \ref{kaon}.
Since a phi meson decays mainly into a kaon-antikaon pair,
medium modifications of kaon properties 
affect the final phi meson yield determined from the
reconstruction of $K^+K^-$ pairs. In the last scenario,
we include both kaon and phi meson medium effects. 
For the in-medium phi meson
mass we use the results from QCD sum rule calculations by Hatsuda and
Lee \cite{hat},

\begin{eqnarray}
{m_\phi ^*\over m_\phi} \approx 1.0-0.0225 {\rho\over \rho_0},
\end{eqnarray}
which is obtained by taking the nucleon strangeness content to be about 0.15.  

\subsection{The phi meson yield}

\begin{figure}[p]
\begin{center}
\vfill
\mbox{\epsfxsize=14truecm\epsffile{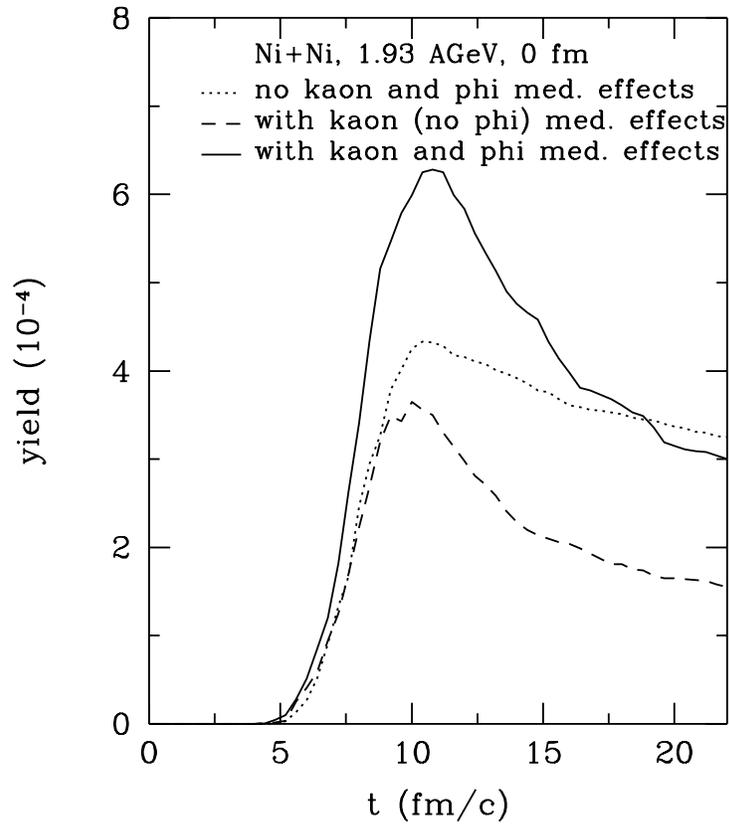}}
\caption{The phi meson yield in Ni+Ni collision 
at 1.93 AGeV and impact parameter of 0 fm for three different scenarios. 
\label{phiabnd}}
\vfill
\end{center}
\end{figure}

The results for phi meson yield in central Ni+Ni collisions
at 1.93 AGeV are shown in Fig. \ref{phiabnd}. 
Phi mesons are mostly produced during the first 10 fm/c, so
the yield increases with time. Afterwards,
phi meson decay and reabsorption become more important, and the phi
meson yield is seen to decrease with time. We find that at
$t\approx 22$ fm/c two-body scattering becomes rather scarce, so
we take this as the freeze-out time and determine the final phi meson yield
by the number of phi mesons which have not decayed. 
                                   
In the scenario that both kaon and phi meson medium effects 
are neglected, the phi meson yield has a value of $3.3 
\times 10^{-4}$, while that of 
$K^-$ is about $1.4\times 10^{-3}$. The  $\phi /K^{-} $ ratio in central 
collision is thus about 25\% and is about a factor of 2 too large 
compared with preliminary FOPI data. However, as pointed out earlier 
\cite{akaon,cass}, the $K^{-}$ yield without medium effects is 
inconsistent with experimental results. In order to explain the 
observed $K^-$ yield, kaon medium effects are needed, and this will 
increase $K^-$ yield by about a factor of 3. 
The $\phi /K^-$ ratio would then be about 8\% and is in 
reasonable agreement with the FOPI data.

On the other hand, the inclusion of kaon medium effects
inevitably affects the phi decay width. With a lower in-medium
$K{\bar K}$ threshold, the phi meson decay width increases significantly
\cite{kkonly}, so the probability for a phi meson to decay inside the
dense matter is enhanced. Because of strong final-state interactions of kaons 
and antikaons, these phi mesons 
cannot be identified from $K^+K^-$ reconstruction. 
Including this effect we find that the 
final phi meson yield is reduced by about a factor of 2 to
about $1.6 \times 10^{-4}$. The $\phi /K^- $ ratio is now about 4\% and 
is about a factor of three too small compared to the FOPI data.

Including also the dropping of phi meson mass in dense
matter leads to an increase of phi meson yield, as a result
of lower production threshold. 
For example, around 2$\rho_0$ where most phi mesons
are produced, the reduction in the production threshold is
about 45 MeV. The final phi meson yield after including both 
kaon and phi meson medium effects is about
$3\times 10^{-4}$, so the $\phi /K^-$ ratio is brought back to about 8\%,
and is again in reasonable agreement with preliminary data from the FOPI
collaboration. 

\begin{figure}[p]
\begin{center}
\vfill
\mbox{\epsfxsize=14truecm\epsffile{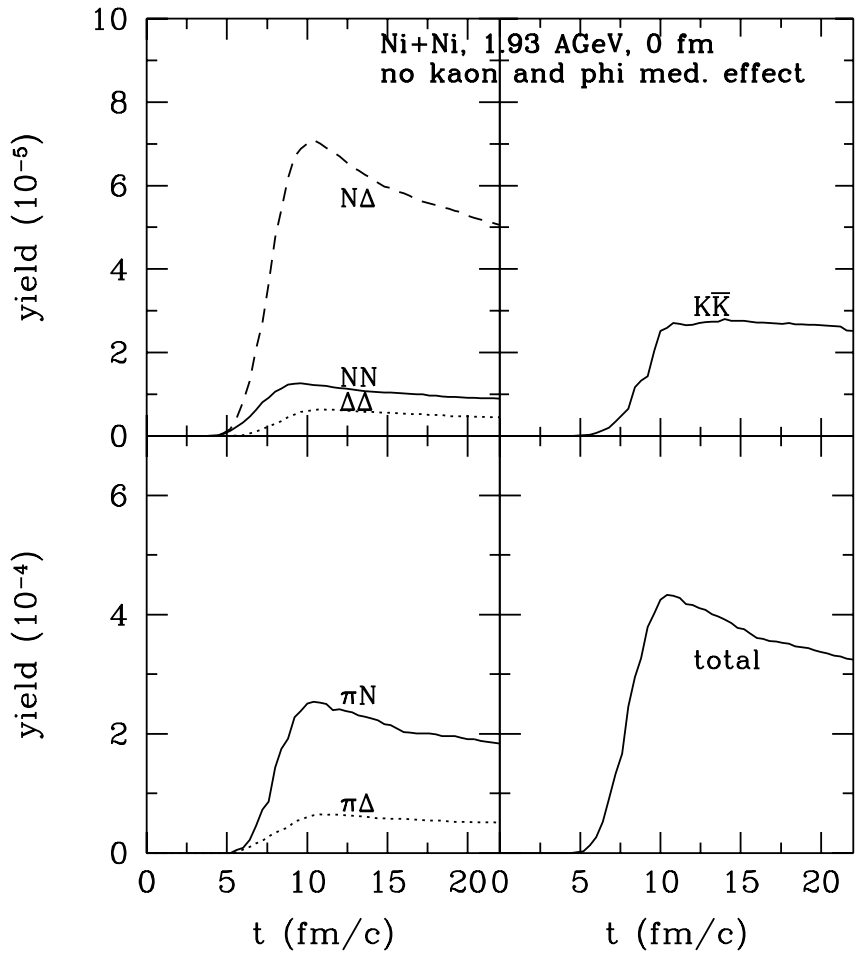}}
\caption{The time evolution of phi meson yield 
from different channels in central Ni+Ni 
collisions at 1.93 AGeV without kaon and phi meson medium effects. 
\label{phipp} }
\vfill
\end{center}
\end{figure}

\begin{figure}[p]
\begin{center}
\vfill
\mbox{\epsfxsize=14truecm\epsffile{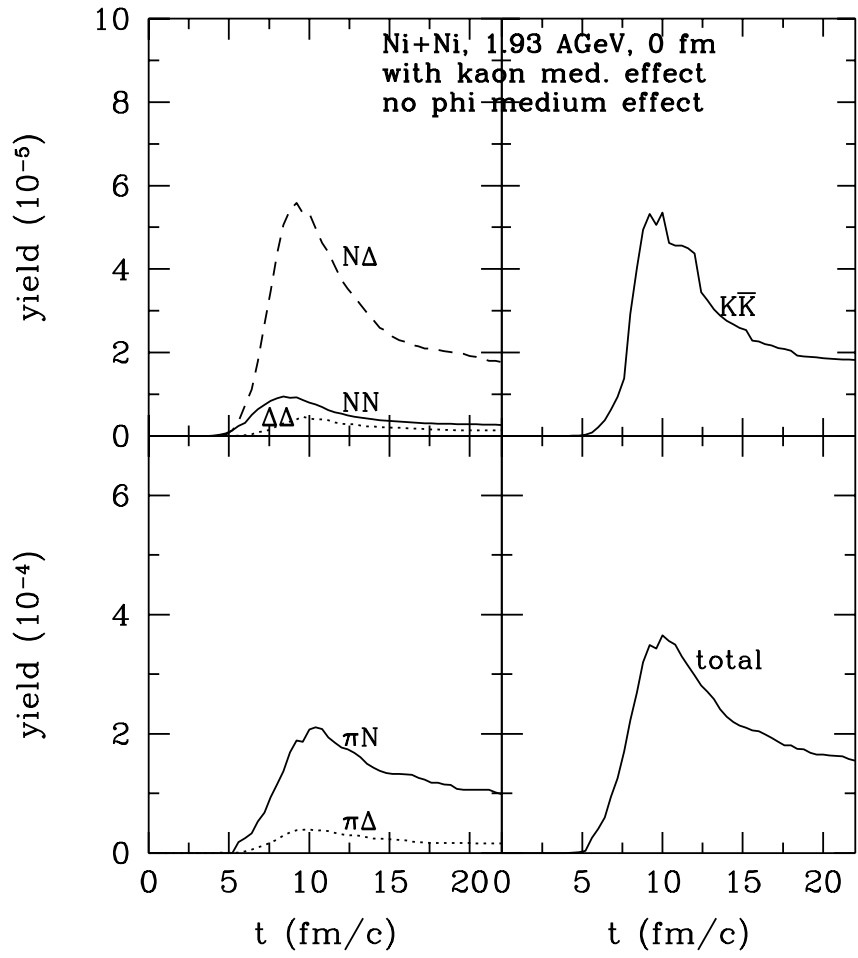}}
\caption{Same as Fig. \protect\ref{phipp} with kaon medium 
effects but no phi meson medium effects. 
\label{phimp} }
\vfill
\end{center}
\end{figure}

\begin{figure}[p]
\begin{center}
\vfill
\mbox{\epsfxsize=14truecm\epsffile{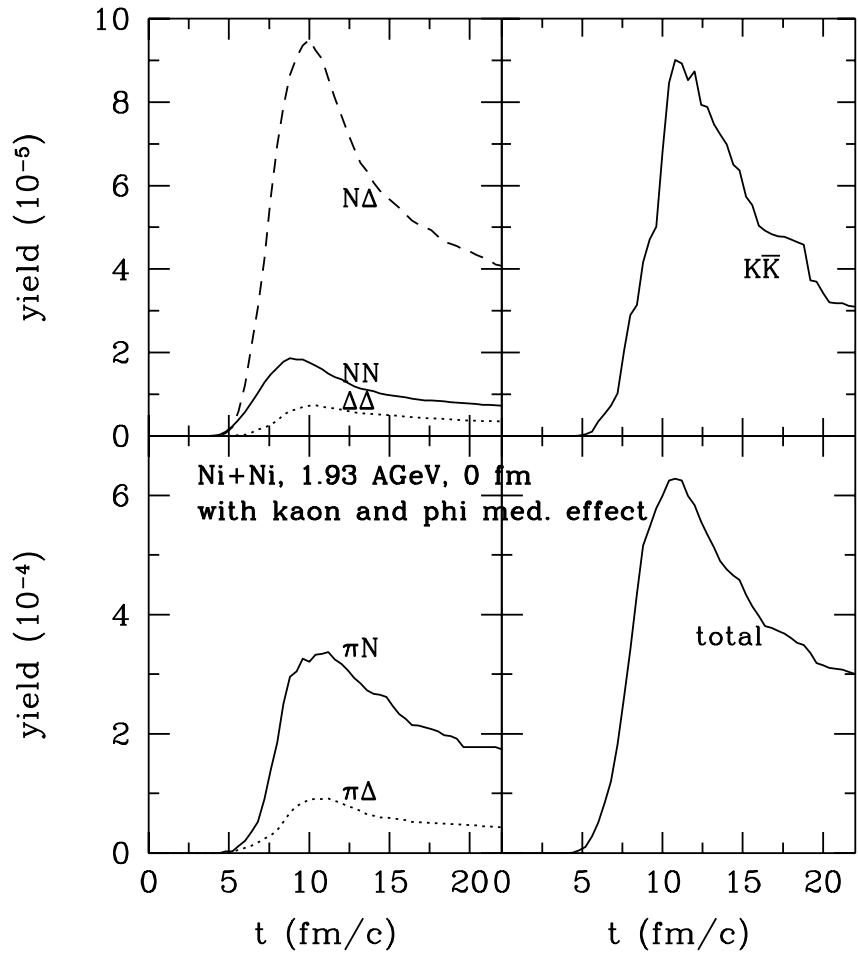}}
\caption{Same as Fig. \protect\ref{phipp} with both
kaon and phi meson medium effects. 
\label{phimm} }
\vfill
\end{center}
\end{figure}

In Figs. \ref{phipp}, \ref{phimp}, and \ref{phimm}, 
the phi meson yield from different production 
channels is shown for the three scenarios mentioned above. 
In all three cases the $\pi N$ channel is the dominant 
one, contributing about 60\% of the total phi meson yield,
The importance of the $\pi N$ channel is due to its large 
cross section near the threshold, although pions 
have a lower abundance than that of nucleons and the 
production threshold for this process is also the highest.
The baryon-baryon channels add up to 
about 15-20\% of the total phi meson yield, with $N \Delta$ 
contributing the most. If phi meson production 
cross sections in $N\Delta$ and $\Delta\Delta$
collisions are assumed to be the same as that in $NN$ collisions, 
we find that the final phi meson yield is reduced
by about 10\%, with the baryon-baryon contribution
being only about 5-10\% of the total.
In the scenario that both kaon and phi meson medium
effects are neglected, the $K \bar K$ channel 
contributes only about 7\% of the total phi meson yield.  
Inclusion of kaon medium effects increases it 
slightly to about 12\% of total phi meson yield. 

\begin{figure}[p]
\begin{center}
\vfill
\mbox{\epsfxsize=14truecm\epsffile{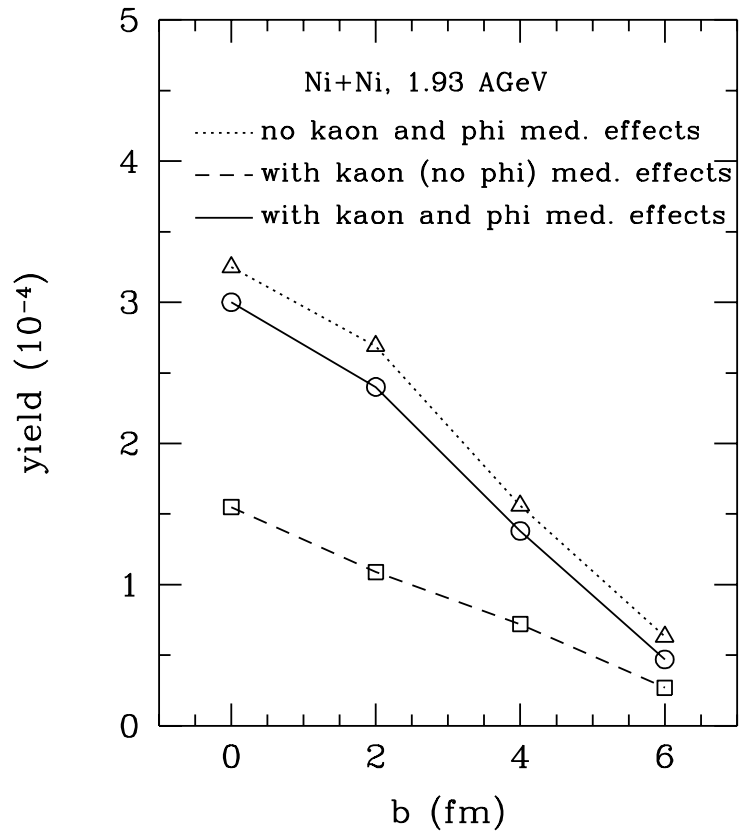}}
\caption{The phi meson yield in Ni+Ni collision 
at 1.93 AGeV at different impact parameters for three different scenarios.
\label{phiimpa}}
\vfill
\end{center}
\end{figure}

The impact parameter dependence of phi meson yield in Ni+Ni
collisions at 1.93 AGeV is shown in Fig. \ref{phiimpa}.
From this, the minimum-biased phi meson production cross section 
can be estimated. It is about 0.15 mb
when both kaon and phi medium effects are neglected.
This reduces to about 0.05 mb if kaon medium effects
are included but those on phi mesons are neglected.
Including both kaon and phi meson medium effects increases 
the production cross section to about 0.13 mb.

Both phi meson decay and absorption have significant effects on its
final yield. Without kaon medium effects, about 20\% of produced phi
mesons are absorbed in $\phi N $ scattering, and about another 
20\% decay before freeze out. Including kaon medium 
effects, absorption by nucleons still removes about 20\% of produced
phi mesons, but phi meson decay is significantly 
enhanced, leading to more than 50\% reduction of the phi meson yield.

\subsection{The phi meson rapidity distribution and transverse 
mass spectrum}

\begin{figure}[p]
\begin{center}
\vfill
\mbox{\epsfxsize=14truecm\epsffile{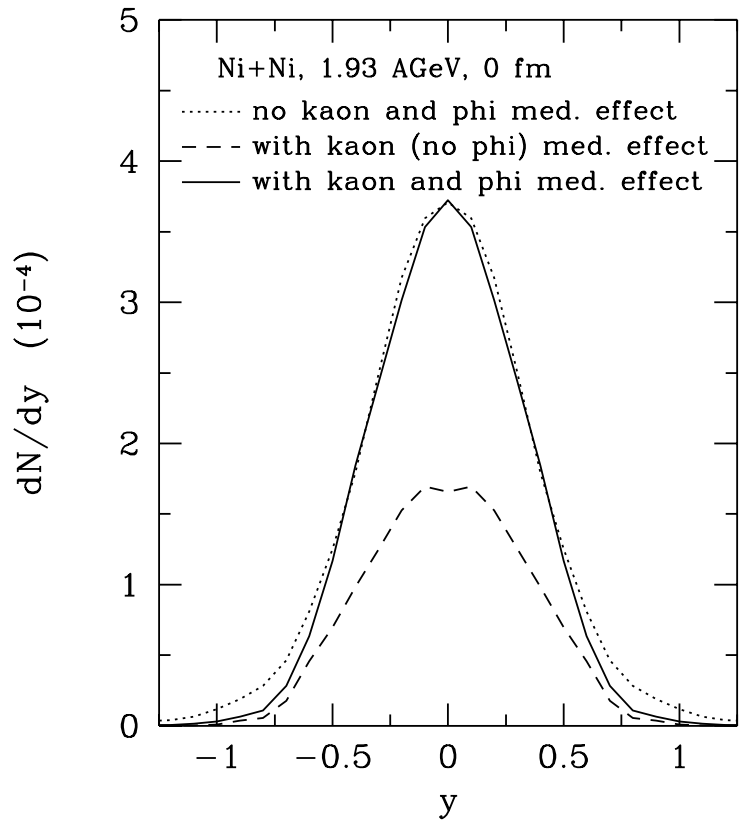}}
\caption{The rapidity distributions of phi 
mesons in central Ni+Ni collisions at 1.93 AGeV for three different 
scenarios. 
\label{phiy} }    
\vfill
\end{center}
\end{figure}

\begin{figure}[p]
\begin{center}
\vfill
\mbox{\epsfxsize=14truecm\epsffile{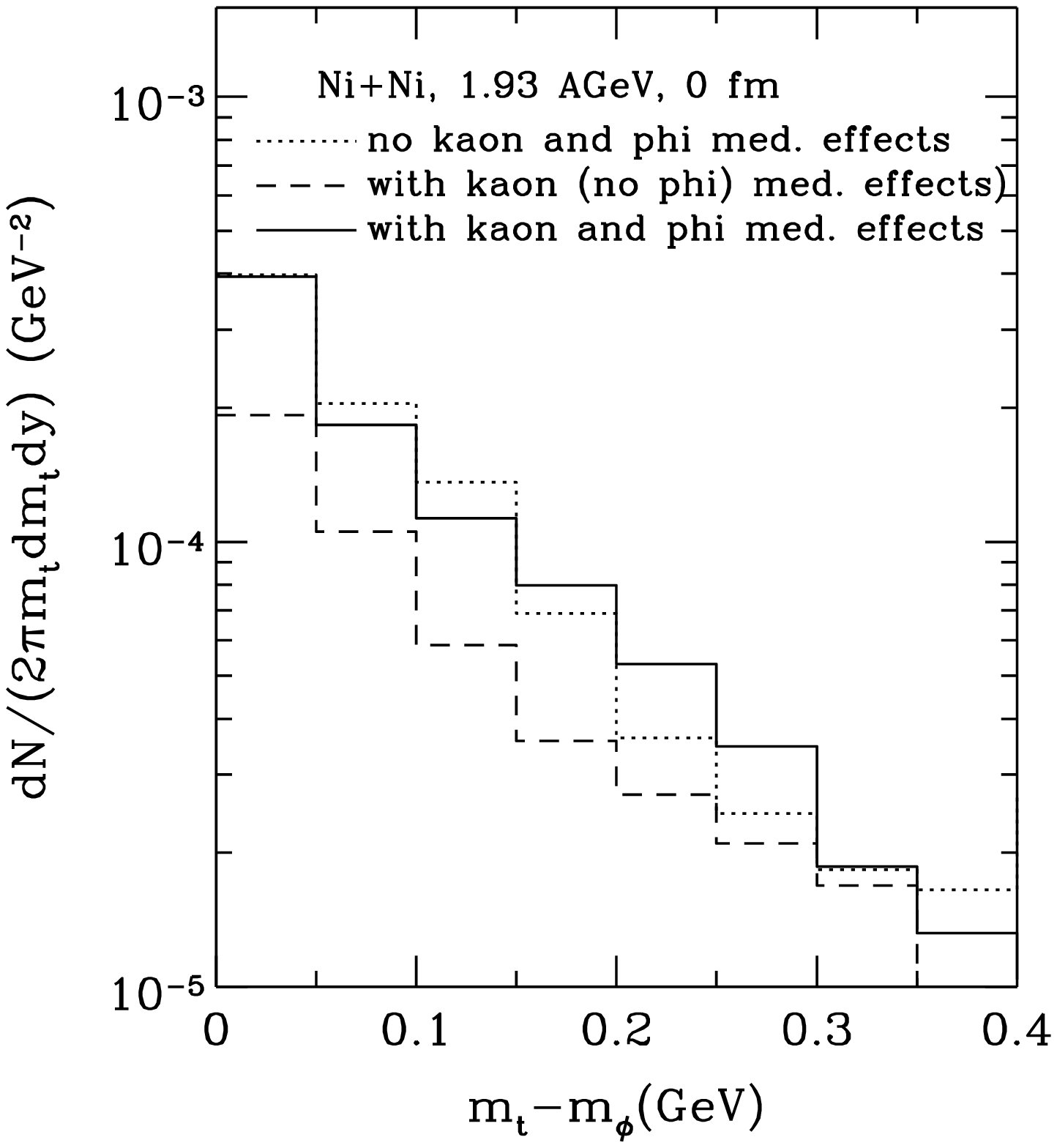}}
\caption{Same as Fig. \protect\ref{phiy} for the transverse mass
distribution. 
\label{phimt} }
\vfill
\end{center}
\end{figure}

The rapidity distribution and transverse mass spectrum
of phi mesons at freeze out are shown in Fig. \ref{phiy}
and Fig. \ref{phimt}, respectively. The phi meson rapidity distribution  
is seen to peak at $y=0$ as those of nucleons, pions, 
and kaons. Their transverse mass distribution has 
a typical exponential form with 
an inverse slope parameter of about 110 MeV for
all three scenarios considered. 

\begin{figure}[p]
\begin{center}
\vfill
\mbox{\epsfxsize=14truecm\epsffile{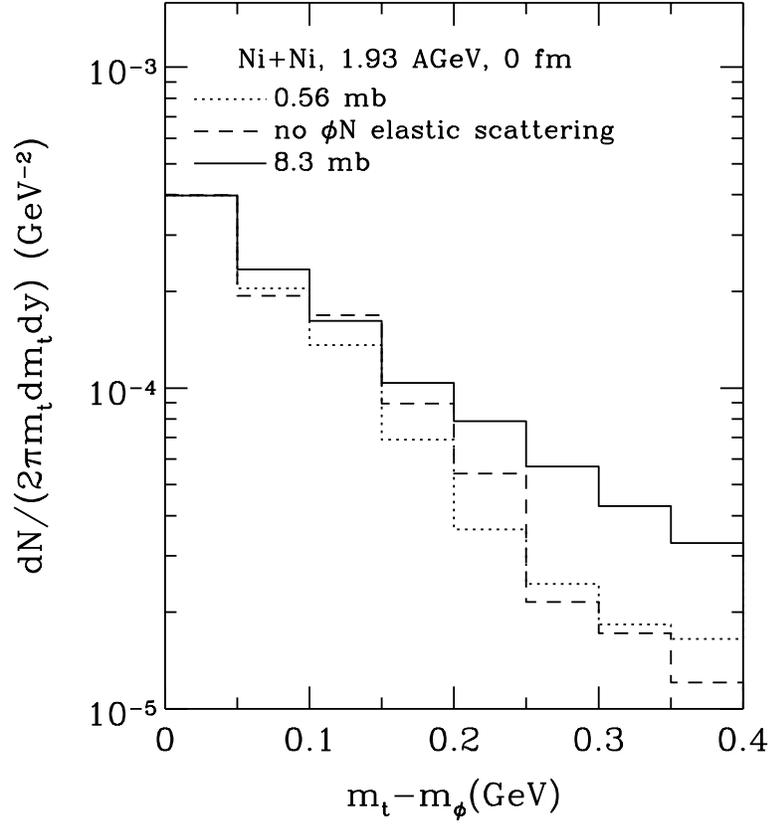}}
\caption{The transverse mass distribution of 
phi mesons in central Ni+Ni collisions at 
1.93 AGeV with (dotted curve for $\sigma_{\phi N}=0.56$ mb and
solid curve for $\sigma_{\phi N}=8.3$ mb) 
and without (dashed curve) $\phi N$ scattering.
\label{phirs}}
\vfill
\end{center}
\end{figure}

The results shown in the above include $\phi N$ scattering with a cross 
section of 0.56 mb. To see more clearly the effects due to scattering 
we show in Fig. \ref{phirs} the phi meson transverse mass spectra 
with and without $\phi N$ elastic scattering in the scenario of
no medium effects on kaons and phi mesons. To facilitate the comparison,
different results are normalized to have the same value at 
$m_t-m_\phi=25$ MeV.  We find that the inverse slope parameter for phi mesons 
in the mid-rapidity region, obtained by fitting the spectrum up to 0.4 GeV, 
changes only slightly when rescattering is included.
As the extraction of $\phi N$ elastic cross section from the experimental 
data is model-dependent, its value may not be well determined. To see 
the effects 
due to a different cross section, we have also repeated the calculation 
using a $\phi N$ elastic cross section of 8.3 mb, which 
is the total $\phi N$ cross section estimated from the 
phi meson photoproduction data \cite{phipho} and 
serves as an upper bound on the elastic cross section. The 
inverse slope parameter in this calculation is found to be 
about 130 MeV. The increase of the slope parameter due to scattering 
can be understood as follows. At SIS energies most 
produced phi mesons have low momenta, so scattering 
with nucleons increases their momenta and helps them in achieving
thermalization with nucleons. Similar effects have been found for 
kaons in heavy ion collisions at subthreshold energies \cite{Fang,aich96}.  

\section{Summary and Outlook}
\label{summary}
We have carried out a transport model study of phi meson production 
from heavy ion collisions 
at SIS energies. The production channels included in our study are $\pinphin$, 
$\pidelnphi$, $\nnnnphi$, $\ndelnnphi$, $\deldelnnphi$, and $\kkphi$.
The cross sections for the first five reactions are obtained from the 
one-boson-exchange model, while that for the last one is taken to be 
of Breit-Wigner form through the phi meson resonance. More than half of
the phi mesons are found to be produced from $\pi N$ collisions.

The predicted $\phi /K^{-}$ ratio depends on whether medium effects are 
included for kaons and phi mesons. Neglecting both kaon and phi medium
effects, the ratio is about 25\%, which is about a factor of two 
larger than that from the FOPI data. However, the $K^-$ yield in this case is 
about a factor of 3 below the measured value. To account for the 
observed $K^-$ yield requires the inclusion of medium effects on 
kaons, such as that predicted by the chiral perturbation theory,
so that the antikaon production threshold is reduced.  
The $\phi /K^-$ ratio turns out to be about a factor 
of 2-3 smaller than the experimental results. 
Including also dropping phi meson mass 
in nuclear medium as predicted by the QCD sum rules increases the phi meson
yield.  The final $\phi/K^-$ ratio becomes about 8\% and agrees 
reasonably with the preliminary FOPI data.

It will be of interest to also measure  
phi mesons through their dilepton decay channel, which will 
be carried out in the near future by the HADES collaboration 
\cite{Hades} at GSI. Since both the $K \bar K$ and 
the dilepton channels are subject to same uncertainties 
in the production mechanism, the ratio of the phi meson yield 
measured from the $K \bar K$ channel to that measured from 
the dilepton channel is expected to be less sensitive to such 
uncertainties. In addition, the 
dilepton channel can also reveal directly the properties 
of vector mesons in nuclear medium, especially the change in 
their masses \cite{Wolf,likobrown,cass95}.  For phi mesons, this 
is not possible via the $K \bar K$ channel since the correlation of a
$K \bar K$ pair is lost due to $K N$ scattering 
in the nuclear medium. Furthermore, the shape 
of the phi meson peak in the dilepton spectrum 
can also indirectly give information 
on the kaon medium effects \cite{shuryak}. For this purpose, a careful study of 
the background is needed to determine whether the phi 
meson peak will still be observable. Previous study \cite{Wolf} 
shows that the background due to nucleon and 
pion bremsstrahlung as well as the eta Dalitz decay
is small near the phi meson peak. 
The major background is thus expected to come from decays of 
rho and omega mesons. A careful study of the production 
of vector mesons in heavy ion collisions is needed in order to learn the phi 
meson properties from the dilepton decay channel. Such a study is in progress.

\begin{center}
\noindent {\bf Acknowledgement}
\end{center}

We acknowledge useful discussions with R. Machleidt. We also thanks 
J. A. M. Vermaseren for his program FORM (version 1.1) 
and P. Lepage for his subroutine VEGAS, which we have used extensively
in evaluating the elementary cross sections for phi meson production. 
This work was
supported in part by the National Science Foundation under
Grant No. PHY-9509266. GQL was also supported in part by
the Department of Energy under Contract  No. DE-FG02-88Er40388.

\pagestyle{empty}

\end{document}